\PassOptionsToPackage{table,dvipsnames}{xcolor}

\documentclass[aps,prb,onecolumn,10pt,showkeys,superscriptaddress]{revtex4-2}
\usepackage[T1]{fontenc}
\usepackage[utf8]{inputenc}
\usepackage{microtype}
\usepackage{newtxtext,newtxmath}
\usepackage{graphicx}
\usepackage{xcolor}          
\usepackage{booktabs}
\usepackage{multirow}
\usepackage{physics}
\usepackage[normalem]{ulem}
\usepackage{cancel}
\usepackage{hyperref}
\hypersetup{
	colorlinks=true,
	linkcolor=blue,
	citecolor=blue,
	filecolor=blue,
	urlcolor=blue,
	breaklinks=true
}
\usepackage{orcidlink}

\definecolor{myMagenta}{RGB}{220,0,110}

\begin{document}

\title{ ModMax-AdS Black Hole with Global Monopole as Source in Kalb-Ramond Gravity}

\author{Faizuddin Ahmed\orcidlink{0000-0003-2196-9622}}
\email{E-mail: faizuddinahmed15@gmail.com}
\affiliation{Department of Physics, The Assam Royal Global University, Guwahati, 781035, Assam, India}

\author{Ahmad Al-Badawi\orcidlink{0000-0002-3127-3453}}
\email{E-mail: ahmadbadawi@ahu.edu.jo}
\affiliation{Department of Physics, Al-Hussein Bin Talal University, 71111, Ma'an, Jordan}

\author{Edilberto O. Silva\orcidlink{0000-0002-0297-5747}}
\email{E-mail: edilberto.silva@ufma.br (Corresponding author)}
\affiliation{Programa de Pós-Graduação em Física \& Coordenação do Curso de Física-Bacharelado,
Universidade Federal do Maranhão, 65085-580 São Luís, Maranhão, Brazil}

\date{\today}

\begin{abstract}
In this work, we investigate in detail the thermodynamic properties of a spherically symmetric ModMax-AdS black hole sourced by a global monopole within the Kalb-Ramond gravity. We derive the key thermodynamic quantities, including the Hawking temperature, Gibbs free energy, and specific heat capacity, and analyze how the geometric parameters influence these physical quantities. The first law of thermodynamics and the corresponding Smarr formula are explicitly verified. Furthermore, we study the thermodynamic criticality of the system by deriving the critical points and examining the effects of the space-time geometric parameters. We also obtain the inversion temperature and demonstrate that the minimum inversion temperature is modified by the space-time parameters. In addition, the sparsity of Hawking radiation and thermal fluctuations of the system are investigated, highlighting the effects of the parameters on the entropy corrections. Finally, we analyze the optical properties of the black hole, in particular the photon sphere and shadow radius, showing how these parameters influence these features.
\end{abstract}

\maketitle

\tableofcontents

\section{Introduction} \label{sec1}

Black holes emerge as a fundamental prediction of Einstein’s theory of general relativity, having been theoretically anticipated nearly a century before their first direct observational confirmation \cite{EHTL1}. Within the Einstein-Maxwell framework, black holes may carry electric charge in addition to mass and angular momentum. A prominent exact solution describing a rotating and electrically charged black hole is the Kerr-Newman spacetime \cite{IDN1989}. More generally, Einstein-Maxwell theory admits solutions endowed with additional parameters, such as the NUT charge and acceleration. Among these, the Plebański-Demiański metric stands out as one of the most general exact solutions, incorporating a wide range of physical parameters within a unified framework \cite{JFP1976}.

In realistic astrophysical environments, black holes are rarely isolated and are often immersed in external magnetic fields. Recent observations by the Event Horizon Telescope have revealed highly organized magnetic field structures near the event horizon of the supermassive black hole Sagittarius~A*, indicating the presence of strong magnetized plasma in its vicinity \cite{EHTL25}. Earlier studies by the same collaboration reported the detection of intense magnetic fields surrounding supermassive black holes at galactic centers, inferred from measurements of polarized radiation \cite{EHTL12,EHTL13}. In particular, the handedness of circularly polarized light propagating near the black hole encodes valuable information about the surrounding magnetic field configuration and the population of high-energy particles. These observational breakthroughs have significantly strengthened the motivation for theoretical models of magnetized black holes and their application to the study of particle dynamics and orbital motion in strong gravitational fields.

Nonlinear electrodynamics (NED) was originally formulated as a generalization of Maxwell's classical theory to overcome certain limitations, most notably the divergence of the electromagnetic field at the location of a point charge. In regimes of strong electromagnetic fields, quantum corrections naturally induce nonlinear modifications to Maxwell's theory, motivating the study of NED in both gravitational and cosmological contexts. In particular, coupling specific NED models to general relativity has been proposed as a possible mechanism for driving cosmic inflation \cite{VADL2002}. Beyond cosmology, NED frameworks have proven useful in addressing spacetime singularities, including those present in black hole solutions \cite{JMB1968,EAB1998}. Among the various NED models, the Born-Infeld (BI) theory has received significant attention due to its remarkable property of yielding finite self energy for point charges and its applications in both string theory and gravitational physics.

Recent advances in modified gravity and nonlinear electrodynamics have opened new avenues for exploring black hole solutions beyond the standard Einstein-Maxwell framework. In this context, ModMax theory-a nonlinear generalization of Maxwell’s electromagnetism-has attracted considerable attention \cite{IB2020,BPK2020,IB2021,PKT2021,DFA2021,ZA2021,ABB2021}. This theory introduces a dimensionless parameter that governs the degree of nonlinearity in the electromagnetic field, smoothly interpolating between the linear Maxwell theory and its nonlinear extension \cite{HMS2024}. When incorporated into black hole solutions, particularly in an anti-de Sitter (AdS) background, ModMax electrodynamics leads to a richer thermodynamic structure and novel phase behavior. The non-linear parameter effectively ``screens" the charges through an exponential factor. In particular, electrically and/or magnetically charged AdS black holes coupled to a ModMax field exhibit significantly modified thermodynamic properties compared to their RN AdS counterparts, including shifts in critical points and stability regions \cite{SIK2022, BES2024,FK2024}. Some recent investigation, include a dyonic Einstein-ModMax black hole \cite{RCP2022}, the study of asymptotically AdS accelerated black holes \cite{JB2022}, stability and pair production of scalars in the charged Einstein-ModMax black hole spacetime \cite{HMS2023}, braneworld black holes with ModMax
electrodynamics \cite{HMS2024b}, ModMax with quintessence field \cite{ALBADAWI2025101865}, aspects of ModMax (A)dS black holes, such as, thermodynamics properties, heat engine, shadow, and photon trajectory \cite{BES2025}, the propagation of light in ModMax spacetime \cite{EGH2024a,EGH2024b}, scattering and absorption of massless scalar waves by a ModMax black hole \cite{AQB2025}, the study of multiblack hole spacetimes in Einstein-ModMax theory \cite{AB2025}, and thermal characteristics of AdS ModMax black hole \cite{MRS2024,GB2025}.

Einstein's general relativity underpins modern cosmology, offering a precise description of the evolution of the Universe. By extending the principles of special relativity to include gravity, GR preserves local Lorentz symmetry at each point in the spacetime manifold. Although experiments and observations strongly support Lorentz symmetry as a fundamental feature of nature, certain theoretical frameworks, particularly string theory, suggest that Lorentz symmetry may be broken at high energy scales \cite{DJG1985,VAK1989}. These considerations motivate the study of possible Lorentz symmetry breaking (LSB) and provide insights into the fundamental structure of spacetime. A general framework for investigating LSB is the Standard-Model Extension, which incorporates gravitational interactions into the Standard Model while allowing for potential violations of Lorentz symmetry \cite{VAK2004}.  

A prominent example of a field that can induce spontaneous LSB is the Kalb–Ramond (KR) field, a rank-two antisymmetric tensor emerging naturally in the spectrum of bosonic string theory \cite{MK1974}. The properties of the KR field have been extensively studied in various gravitational and cosmological contexts \cite{WFK1996,SK2003,KKN2022}. In particular, a non-minimal coupling of the KR field to gravity provides a mechanism for spontaneous LSB: when the KR field acquires a nonzero vacuum expectation value (VEV), Lorentz symmetry is broken \cite{BA2010}. Black hole solutions in this framework have been explored in several works. A Schwarzschild-like solution was first derived in \cite{LAL2020}, followed by studies of neutral, static, and spherically symmetric black holes \cite{ref1}, electrically charged black holes \cite{ref2}, and solutions incorporating additional features such as a global monopole \cite{ref3,MF2025} or a cloud of strings \cite{FA2025a}, and ModMax black holes in phantom-enhanced KR gravity \cite{PDU2025}. These studies illustrate the rich structure and phenomenology of black holes in KR-gravity. Related studies that incorporate exotic matter, nonlinear electrodynamics, and dark components have analyzed geodesic properties, black hole shadows, greybody factors, and thermodynamic stability, offering useful comparisons for this work \cite{FA3,FA4,FA1,FA2,FA5,FA6}.

Bekenstein and Hawking identified a deep analogy between the geometric properties of black holes and thermodynamic variables, enhancing our understanding of the connection between gravity and classical thermodynamics \cite{Bekenstein1973,Hawking1974}. Bekenstein introduced the concept of black hole entropy and quantified it through the area law, \(S = \frac{A}{4}.\) However, there are notable differences between black hole thermodynamics and conventional thermodynamics. For instance, black hole entropy is proportional to the horizon area rather than the volume, and the heat capacity of certain black holes can be negative.  

The thermodynamic properties of black holes, particularly phase transitions and thermal stability, provide important insights into the fundamental structure of spacetime geometry. These properties become even more interesting in the context of anti-de Sitter (AdS) black holes. One key reason is the work of Hawking and Page, who discovered a first-order phase transition between the Schwarzschild-AdS black hole and thermal AdS space \cite{Hawking1983}. This phenomenon, known as the Hawking-Page phase transition, has been interpreted as the gravitational dual of the QCD confinement/deconfinement transition \cite{Witten1998a,Witten1998b}. Understanding phase transitions requires a detailed analysis of temperature diagrams that depict the system's states below, at, and above the critical point. These diagrams are often complemented by an examination of the free energy, expressed in either the Gibbs or Helmholtz form, to interpret the corresponding phase behavior. Such analyses typically reveal characteristic patterns: first-order phase transitions appear as swallowtail structures in the free energy curves, whereas second-order phase transitions are indicated by smooth, continuously varying curves without discontinuities. Another significant motivation for studying the thermodynamics of AdS black holes comes from the discovery of phase transitions analogous to the liquid/gas transitions of Van der Waals fluids in Reissner-Nordström-AdS (RN-AdS) black holes by Chamblin \textit{et al.} \cite{Chamblin1999a,Chamblin1999b}. They investigated RN-AdS black holes in the canonical ensemble and identified a first-order phase transition between small and large black holes.  

Interest in AdS black hole thermodynamics has also been driven by treating the cosmological constant as a thermodynamic variable. When the cosmological constant varies, the first law of black hole thermodynamics becomes consistent with the Smarr relation. In this extended framework, a \(V dP\) term is included, where the pressure in Einstein gravity is defined as \(P = -\Lambda/(8\pi),\) and the black hole mass \(M\) is interpreted as the enthalpy of the system rather than its internal energy \cite{Kastor2010,Dolan2011}. One of the first studies in this extended phase space approach was by Kubiznak et al., who demonstrated the existence of a critical behavior in the phase space of charged AdS black holes analogous to the Van der Waals fluid \cite{Kubiznak2012}.  Recent developments in black hole thermodynamics include the study of Joule-Thomson (JT) expansion \cite{Johnson2016,Okcu2017,Okcu2018} and the formulation of holographic heat engines \cite{Ghaffarnejad2018,Zhang2019}. The JT expansion in AdS black holes is inspired by the analogy between AdS black holes and Van der Waals fluids. It occurs at constant enthalpy (equivalent to the black hole mass) and is used to analyze thermal expansion, identifying regimes of heating and cooling. In this context, the inversion temperature and the corresponding inversion pressure mark the transition between heating and cooling during the expansion process.

Black holes emit subatomic particles through Hawking radiation, providing insights into the underlying black hole geometry. Modifying the Bekenstein entropy relation becomes necessary when incorporating concepts such as the holographic principle \cite{Easther1999} and thermal fluctuations. In statistical mechanics, quantum fluctuations manifest as thermal fluctuations around equilibrium, which lead to corrections in the maximal entropy of black holes as their size decreases due to Hawking radiation. At the first-order approximation, these corrections appear as logarithmic terms in the entropy, arising from small statistical perturbations near equilibrium. Quantum fluctuations in the black hole geometry thus induce thermal fluctuations in its thermodynamic behavior. Incorporating these corrections into black hole thermodynamics impacts local and global stability, critical phenomena, holographic duality, and other essential properties. Thermal fluctuation corrections therefore play a crucial role in connecting black hole thermodynamics with the underlying spacetime geometry. As a consequence of thermal fluctuations \cite{Moore2005}, the standard area–entropy relation requires modification by a logarithmic factor. The procedure for deriving this corrected entropy at equilibrium was systematically studied by Das \textit{et al.} \cite{Das2002}. It has been shown that the logarithmic correction is meaningful only when a Schwarzschild black hole is placed within a cavity of sufficiently small radius \cite{Akbar2004}, and it is not applicable when the black hole is thermally unstable and evaporates completely via Hawking radiation. The effects of thermal fluctuations on the entropy of both neutral and charged black holes were further investigated by Gour and Medved \cite{Gour2003}, with subsequent studies extending these results. One approach to deriving the logarithmic correction term is through the density of states method, as discussed in canonical approaches to non-perturbative quantum gravity \cite{Abhay1991}.

In this work, we investigate ModMax black hole solutions with a global monopole in the framework of KR gravity, considering the presence of a cosmological constant. We analyze the effects of various corrections, including the global monopole, Lorentz symmetry breaking, and the ModMax parameter, on the thermodynamic properties of the black hole. In particular, we compute the Hawking temperature and specific heat, and examine thermodynamic criticality by determining the critical points. Moreover, we study the sparsity of Hawking radiation, a quantum effect in which black holes emit thermal radiation due to particle-antiparticle pair creation near the event horizon. Finally, we investigate the optical properties of the black hole by calculating the photon sphere and shadow radius, highlighting the combined influence of the spacetime geometry parameters.

\section{ModMax-AdS BH with GM in KR-gravity} \label{sec2}

In this section, we present a detailed analysis of the static, spherically symmetric ModMax-AdS black hole solution sourced by a global monopole in KR gravity. We focus on both the thermodynamic properties and the optical characteristics, highlighting how various geometric and physical parameters influence the spacetime curvature and consequently affect the thermal behavior and photon properties around the black hole.

In Ref.~\cite{ref1}, the authors presented exact solutions for static and spherically symmetric neutral black holes in the absence and presence of cosmological constant within KR-gravity. Building on this, in Ref.~\cite{ref2}, the authors obtained electrically charged black hole solutions, both with and without a cosmological constant in the same gravity theory. The spacetime geometry describing an electrically charged black hole in this framework is given by the following static and spherically symmetric metric:
\begin{align}
ds^2 &= -h(r)\,dt^2 +\frac{dr^2}{h(r)}+r^2\,d\Omega^2,\qquad d\Omega^2=d\theta^2+\sin^2 \theta d\phi^2,\label{aa1}
\end{align}
where the lapse function is of the following form:
\begin{equation}
h(r)=\frac{1}{1 - \ell} - \frac{2\,M}{r} +\frac{Q^2}{(1 - \ell)^2\,r^2}-\frac{\Lambda}{3\,(1-\ell)}\,r^2.\label{aa2}
\end{equation}
Here $\ell=\xi b^2/2$ is the KR-field parameter, referred to as the Lorentz-violating parameter, measures the magnitude of Lorentz-breaking effects. Here $\xi$ is the coupling constant, $b_{\mu\nu} b^{\mu\nu}=\mp\,b^2$, $Q$ is the electric charge, $M$ is the BH mass, and $\Lambda$ is the usual cosmological constant. 

In Ref. \cite{ref3}, the authors investigated black hole solutions in the same KR-gravity sourced by a global monopole both in the absence and presence of the cosmological constant. For a static and spherically symmetric metric (\ref{aa1}), the lapse function is given by the following form:
\begin{align}
h(r)=\frac{1-k\,\eta^2}{1-\ell}-\frac{2\,M}{r}-\frac{\Lambda_\text{eff}}{3}\,r^2,\label{aa3}
\end{align}
Here $\eta$ being the energy scale of the symmetry-breaking \cite{Barriola1989} and $\kappa=8\pi G$ with $G$ is the Newtonian gravitational constant.

The action for the theory, which includes gravity non-minimally coupled to a self-interacting KR field and ModMax nonlinear electrodynamics, is given by
\begin{align}
S &= \frac{1}{2} \int d^4x \sqrt{-g} \,(R - 2 \Lambda)
+ \int d^4x \sqrt{-g} \Big[
    - \frac{1}{12} H^{\mu\nu\rho} H_{\mu\nu\rho}- V(B^{\mu\nu} B_{\mu\nu} \pm b^2)+ \frac{1}{2} \Big(
    \xi_2 B^{\rho\mu} B^\nu{}_\mu R_{\rho\nu}+ \xi_3 B^{\mu\nu} B_{\mu\nu} R\Big)\Big]\nonumber\\
    &\quad + \int d^4x \sqrt{-g} \, \mathscr{L}_\text{M},
    \label{aa4}
\end{align}
where $\xi_{2,3}$ are the non-minimal coupling constants between gravity and the KR field, $B_{\mu\nu}$ is  a rank-two antisymmetric tensor field representing the KR-field, $H_{\mu \nu \rho }\equiv \partial _{[\mu }B_{\nu \rho ]}$ is field strength. To achieve the charged solutions, we consider the matter Lagrangian is given by:
\begin{equation}
    \mathcal{L}_M=\mathcal{L}_{\rm ModMax}+\mathcal{L}_{\rm int}+\mathcal{L}_{GM},\label{aa5}
\end{equation}
where $\mathcal{L}_{GM}$ is the Lagrangian density for the global monopole \cite{Barriola1989} and others are as follows \cite{IB2020,BPK2020,BES2026}:
\begin{align}
\mathcal{L}_{\rm ModMax}=-\frac{1}{2} (\cos \gamma+\sin \gamma)\,F_{\mu\nu}F^{\mu\nu},\qquad \mathcal{L}_{\rm int}=-\zeta B^{\alpha \beta }B^{\gamma \rho }F_{\alpha \beta }F_{\gamma \rho } \label{aa6}
\end{align}
with $\mathcal{L}_{\rm int}$ represents the interaction between the electromagnetic field and the KR field.

The modified Einstein equations is obtained by varying the action (\ref{aa4}) with respect to the metric $g^{\mu\nu}$ is given by
\begin{equation}
    R_{\mu \nu }-\frac{1}{2}g_{\mu \nu }R+\Lambda g_{\mu \nu }= \kappa \left(T^{\text{ GM }}_{\mu \nu }+T^{\text{ KR }}_{\mu \nu }\right)+T^{\text{ M }}_{\mu \nu },\quad \label{aa7}
\end{equation}
where $T^{\text{ KR }}_{\mu \nu }$ is given in \cite{ref2} and 
\begin{align}
    T^{\text{ M }}_{\mu \nu }=2 e^{-\gamma} F_{\mu \alpha } F_{\nu }{}^{\alpha }- \frac{1}{2} g_{\mu \nu } e^{-\gamma} F^{\alpha \beta } F_{\alpha \beta } +\zeta \left( 8 B^{\alpha \beta } B_{\nu }{}^{\gamma } e^{-\gamma} F_{\alpha \beta } F_{\mu \gamma }-g_{\mu \nu } B^{\alpha \beta } B^{\gamma \rho } e^{-\gamma} F_{\alpha \beta } F_{\gamma \rho } \right),
\end{align}

Furthermore, the modified Maxwell equation is derived by varying the action (\ref{aa4}) with respect to the vector potential $A^{\mu}$  yielding
\begin{equation}
\nabla ^{\nu }\left( e^{-\gamma}\,F_{\mu \nu } + 2\zeta B_{\mu \nu }B^{\alpha \beta } e^{-\gamma}\, F_{\alpha \beta }\right) =0.
\end{equation}

Following the methodology and steps done in \cite{ref1,ref2,ref3}, we present a spherically symmetric ModMax-AdS black hole solution within the KR gravity framework with global monopole as a source. The corresponding static and spherically symmetric spacetime is described by the following line element:
\begin{equation}
ds^2 = -f(r)\,dt^2 +\frac{1}{f(r)}\,dr^2+r^2\,d\Omega^2, \label{metric}
\end{equation}
where the metric function $f(r)$ is given by
\begin{align}
   f(r)=\frac{1-k \eta^2}{1 - \ell} - \frac{2\,M}{r} +\frac{e^{-\gamma}\,Q^2}{(1 - \ell)^2\,r^2}-\frac{\Lambda}{3\,(1-\ell)}\,r^2.\label{function}
\end{align}

Here, $\Lambda < 0$ indicates a negative cosmological constant, which characterizes an AdS background geometry. Due to the presence of the KR field, the effective cosmological constant is given by $\Lambda_{\mathrm{eff}} = \frac{\Lambda}{1 - \ell}$ and the effective electric charge $Q_{\rm eff}=\frac{Q}{1-\ell}$. This modification demonstrates how LV effects can significantly alter the asymptotic behavior of AdS spacetimes, fundamentally changing the large-distance properties of the gravitational field.

\begin{figure}[htb!]
\centering
\includegraphics[width=\textwidth]{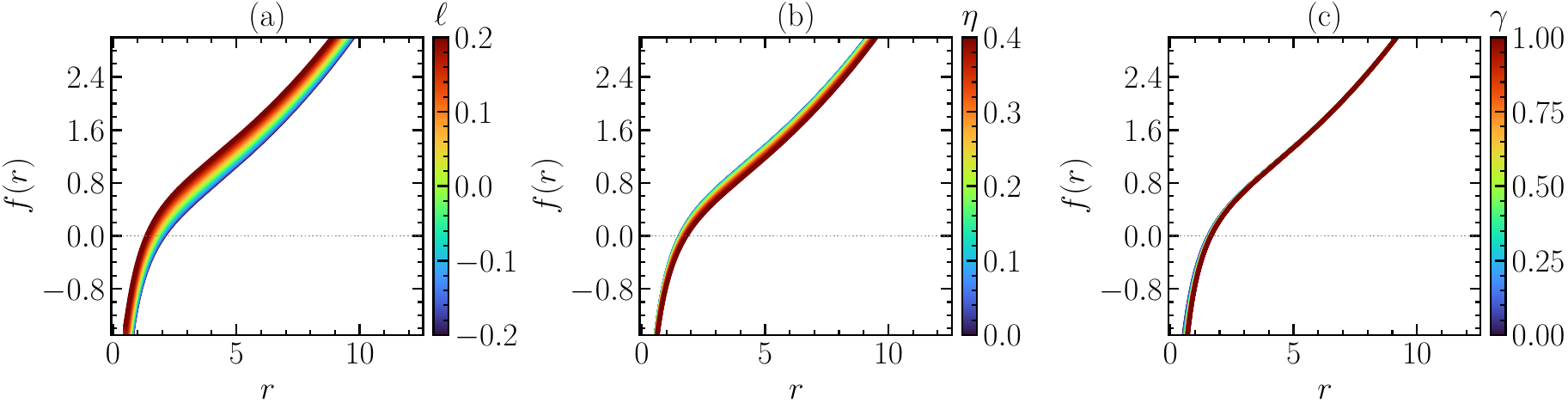}
\caption{Radial profile of the metric function $f(r)$ [Eq.~(\ref{function})] for $M=1$, $Q=0.5$, $k=1$, and $P=0.003$. In each panel, one parameter is varied continuously over a prescribed interval and encoded by the color gradient (see the vertical colorbar), while the remaining parameters are fixed. Panel~(a): $\ell$ varied continuously (with $\eta=0.1$ and $\gamma=0.1$ fixed). Panel~(b): $\eta$ varied continuously (with $\ell=0.1$ and $\gamma=0.1$ fixed). Panel~(c): $\gamma$ varied continuously (with $\ell=0.1$ and $\eta=0.1$ fixed). The gray dotted horizontal line indicates $f(r)=0$.}
\label{fig:metric}
\end{figure}

In Fig.~\ref{fig:metric}, we display the radial profile of the metric function $f(r)$ given in Eq.~(\ref{aa5}) for representative values of the model parameters, with the black hole mass set to $M=1$, the electric charge to $Q=0.5$, the monopole coupling to $k=1$, and the thermodynamic pressure to $P=0.003$. Each panel uses a continuous color gradient (with a vertical colorbar) to encode the parameter being varied within a prescribed interval, which allows one to track the systematic deformation of $f(r)$ without introducing multiple discrete legends. Panel~(a) illustrates the effect of the Lorentz-violating parameter $\ell$ varied continuously; as $\ell$ increases, the asymptotic value $f(r\to\infty)$ is raised due to the rescaling of $\Lambda_{\mathrm{eff}}=\Lambda/(1-\ell)$, and the event-horizon location (given by $f(r_h)=0$) shifts. Panel~(b) shows the influence of the global monopole parameter $\eta$, which primarily shifts the overall level of $f(r)$ via the deficit solid angle term while leaving the near-horizon behavior comparatively less sensitive. Finally, panel~(c) depicts the role of the ModMax parameter $\gamma$: since the electromagnetic contribution enters through $e^{-\gamma}Q^2$, increasing $\gamma$ suppresses the charge sector and modifies the near-horizon structure more strongly than the large-$r$ regime. Overall, Fig.~\ref{fig:metric} highlights that $\ell$ predominantly controls the asymptotics, $\eta$ governs the solid-angle deficit, and $\gamma$ modulates the effective electromagnetic strength near the black hole.

\section{Thermodynamics}\label{sec:2}

The thermodynamic properties of AdS black holes in KR gravity exhibit significant deviations from their general relativistic counterparts due to the non-minimal coupling between the spacetime metric and the antisymmetric KR tensor field. Although the basic thermodynamic framework continues to be governed by the classical laws of black hole mechanics \cite{Bekestein973} and quantum effects still attribute Hawking radiation and entropy to black holes \cite{HawkingPage}, the presence of the KR field introduces nontrivial corrections to key thermodynamic quantities such as the Hawking temperature, entropy, and heat capacity. These corrections can modify the local and global stability conditions, shift the location of phase transition points, and give rise to a richer thermodynamic phase structure compared to the Schwarzschild and RN black holes.

To examine the dependence of thermodynamic state variables on the black hole parameters, we first compute the black hole mass, which is interpreted as the enthalpy of the system in the extended phase space formalism. Using the horizon condition $f(r_h)=0$, where $r_h$ denotes the radius of the event horizon, we obtain
\begin{equation}
 M=\frac{r_h}{2}\left[\frac{1-k \eta^2}{1 - \ell}+e^{-\gamma}\,\frac{Q^2}{(1 - \ell)^2\,r^2_h}-\frac{\Lambda}{3\,(1-\ell)}\,r^2_h\right].\label{dd1}
\end{equation}
In the extended phase space thermodynamics of KR gravity, the cosmological constant $\Lambda$ is identified with the thermodynamic pressure $P$ through the relation $\Lambda=-8\pi P (1-\ell)$. Consequently, the black hole mass can be rewritten as
\begin{equation}
 M=\frac{r_h}{2}\left [\frac{1-k \eta^2}{1 - \ell}+e^{-\gamma}\,\frac{Q^2}{(1 - \ell)^2\,r^2_h}+\frac{8\pi P}{3}\,r^2_h\right].\label{dd2}
\end{equation}

\begin{figure}[htb!]
\centering
\includegraphics[width=\textwidth]{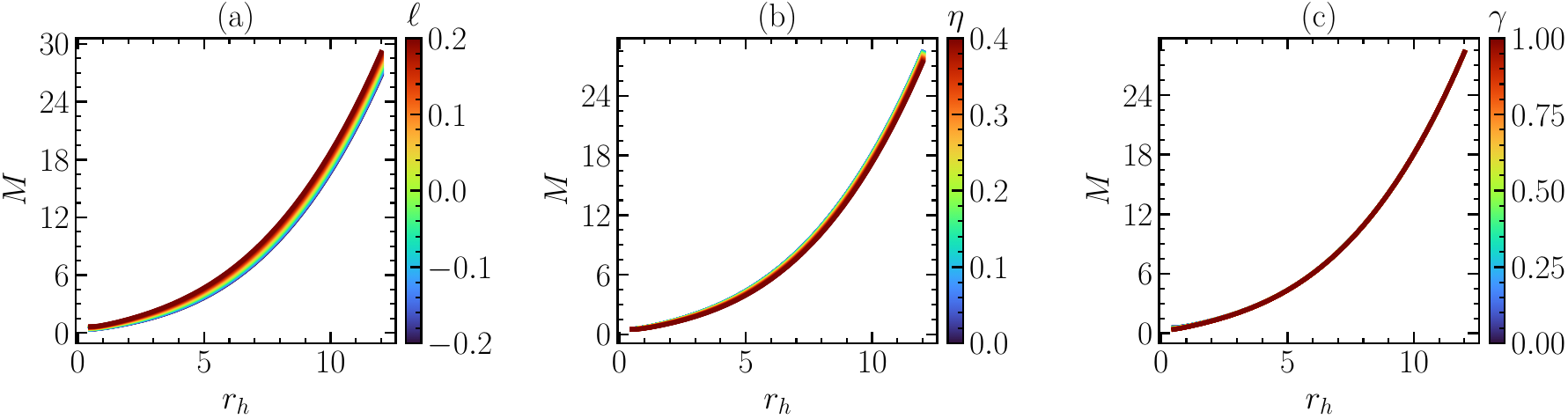}
\caption{Black hole mass $M$ as a function of the event-horizon radius $r_h$ [Eq.~(\ref{dd2})], for $Q=0.5$, $k=1$, and $P=0.003$. In each panel, one parameter is varied continuously and encoded by the color gradient (see colorbar), while the remaining parameters are fixed. Panel~(a): $\ell$ varied continuously (with $\eta=0.1$ and $\gamma=0.1$ fixed). Panel~(b): $\eta$ varied continuously (with $\ell=0.1$ and $\gamma=0.1$ fixed). Panel~(c): $\gamma$ varied continuously (with $\ell=0.1$ and $\eta=0.1$ fixed).}
\label{fig:mass}
\end{figure}

The behavior of the black hole mass as a function of the horizon radius is presented in Fig.~\ref{fig:mass}. In all panels, $M(r_h)$ exhibits a characteristic minimum at a certain radius that separates two physically distinct branches: a small black hole branch at small $r_h$, dominated by the Coulomb-like $Q^2/r_h$ term, and a large black hole branch at large $r_h$, dominated by the pressure term $\propto P\,r_h^3$. Here, the continuously varying parameter in each panel is encoded by a color gradient (see the corresponding colorbar), which makes the systematic displacement of the minimum and the global shift of $M(r_h)$ visually transparent. Panel~(a) shows that increasing $\ell$ raises the mass function at essentially all radii, reflecting the enhancement of the factor $(1-k\eta^2)/(1-\ell)$. Panel~(b) reveals that larger $\eta$ reduces $M(r_h)$, consistent with the deficit solid angle diminishing the effective gravitational mass. Panel~(c) indicates that increasing $\gamma$ slightly lowers $M(r_h)$ because the electromagnetic contribution $e^{-\gamma}Q^2$ is suppressed. The location and depth of the minimum (extremal configuration) therefore depend sensitively on the combined effects of Lorentz violation, monopole charge, and nonlinear electrodynamics.

The surface gravity of the black hole, calculated using Wald's formalism, is given by \cite{Wald1984}
\begin{equation}
\kappa=-\frac{1}{2}\,\lim_{r \to r_h} \frac{\partial_r g_{tt}}{\sqrt{-g_{tt}\,g_{rr}}}=\frac{f'(r_h)}{2}.\label{dd3}
\end{equation}
Accordingly, the Hawking temperature associated with the event horizon is obtained as
\begin{equation}
 T=\frac{\kappa}{2\pi}=\frac{1}{4\pi r_h}\left[\frac{1-k \eta^2}{1-\ell}-e^{-\gamma}\,\frac{Q^2}{(1-\ell)^2 r^2_h}+8\pi P\,r^2_h\right].\label{dd4}
\end{equation}

\begin{figure}[htb!]
\centering
\includegraphics[width=\textwidth]{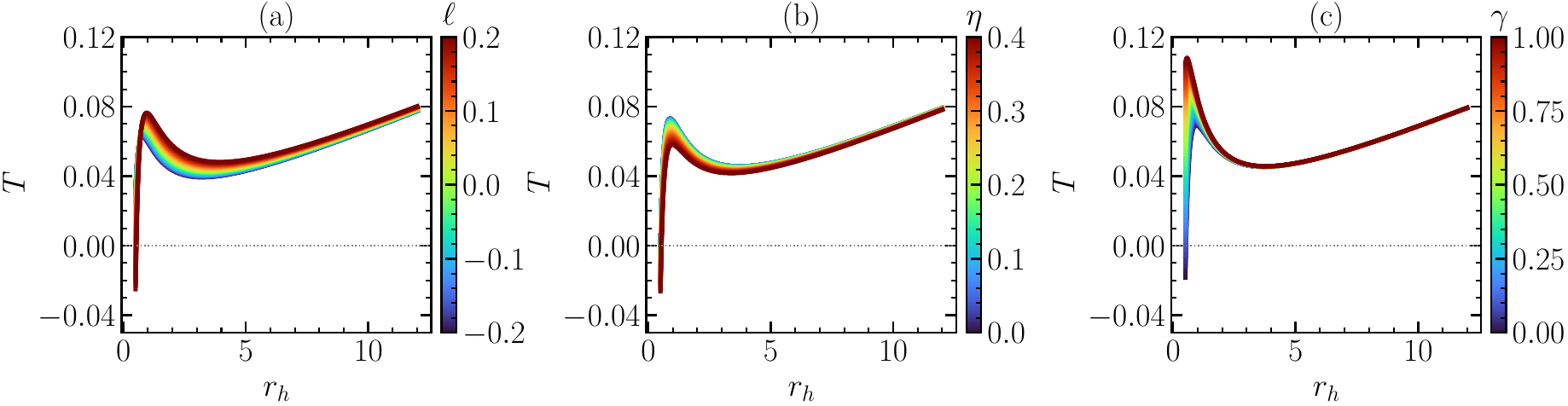}
\caption{Hawking temperature $T$ as a function of the event-horizon radius $r_h$ [Eq.~(\ref{dd4})], for $Q=0.5$, $k=1$, and $P=0.003$. In each panel, one parameter is varied continuously and encoded by the color gradient (see colorbar), while the remaining parameters are fixed. Panel~(a): $\ell$ varied continuously (with $\eta=0.1$ and $\gamma=0.1$ fixed). Panel~(b): $\eta$ varied continuously (with $\ell=0.1$ and $\gamma=0.1$ fixed). Panel~(c): $\gamma$ varied continuously (with $\ell=0.1$ and $\eta=0.1$ fixed). The gray dotted horizontal line indicates $T=0$.}
\label{fig:temperature}
\end{figure}

The Hawking temperature as a function of the horizon radius is plotted in Fig.~\ref{fig:temperature}. The temperature profile displays a structure typical of charged-AdS black holes: at small $r_h$, $T$ is dominated by the negative charge term, which can render $T<0$ and thus excludes very small horizons; at larger $r_h$, the pressure term drives $T$ upward. The continuous color encoding in each panel highlights how the location of the extrema (when present) and the overall temperature scale evolve smoothly with the model parameters. In panel~(a), increasing $\ell$ shifts the temperature profile upward by amplifying the prefactor $(1-\ell)^{-1}$. In panel~(b), increasing $\eta$ also raises the curve through the factor $(1-k\eta^2)$, although the overall effect is typically milder in the explored interval. In panel~(c), larger $\gamma$ weakens the electromagnetic sector (via $e^{-\gamma}$), which tends to reduce the magnitude of the negative charge contribution at small radii and thereby reshapes the low-$r_h$ part of the curve, while leaving the large-$r_h$ pressure-dominated regime comparatively unchanged.

Furthermore, the black hole entropy is given by the Bekenstein-Hawking formula \cite{Bekestein973, Gibbons1977}, which states that it is equal to one-quarter of the event horizon area. This can be derived via the integral
\begin{equation}
S = \int \frac{dM}{T_H} = \pi r_h^2 = \frac{A}{4},\label{dd5}
\end{equation}
where $A $ is the horizon area.

We now proceed to evaluate the Gibbs free energy of the black hole, which plays a crucial role in analyzing the global thermodynamic stability and phase structure of the system. The Gibbs free energy is defined in terms of the black hole mass, temperature, and entropy as
\begin{align}
G= M - T S=\frac{r_h}{4}\left[\frac{1-k \eta^2}{1-\ell}+3\frac{e^{-\gamma}\,Q^2}{(1-\ell)^2 r^2_h}-\frac{8\pi P}{3}\,r^2_h\right].\label{dd6}
\end{align}

\begin{figure}[htb!]
\centering
\includegraphics[width=\textwidth]{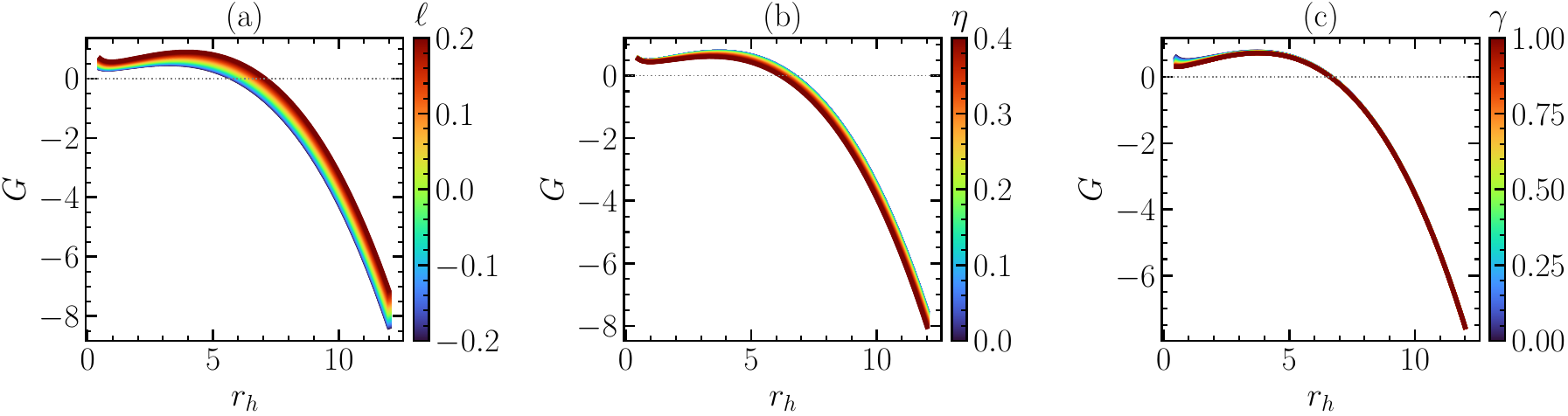}
\caption{Gibbs free energy $G$ as a function of the event-horizon radius $r_h$ [Eq.~(\ref{dd6})], for $Q=0.5$, $k=1$, and $P=0.003$. In each panel, one parameter is varied continuously and encoded by the color gradient (see colorbar), while the remaining parameters are fixed. Panel~(a): $\ell$ varied continuously (with $\eta=0.1$ and $\gamma=0.1$ fixed). Panel~(b): $\eta$ varied continuously (with $\ell=0.1$ and $\gamma=0.1$ fixed). Panel~(c): $\gamma$ varied continuously (with $\ell=0.1$ and $\eta=0.1$ fixed). The gray dotted horizontal line indicates $G=0$.}
\label{fig:gibbs}
\end{figure}

The Gibbs free energy $G(r_h)$ is depicted in Fig.~\ref{fig:gibbs}. In all panels, $G$ decreases as $r_h$ grows and crosses $G=0$ at a parameter-dependent radius, signaling the Hawking--Page transition between thermal AdS and the black hole phase. With the continuous color encoding, one can directly read off how the zero-crossing and the overall level of $G(r_h)$ shift as a function of each parameter. Panel~(a) shows that increasing $\ell$ elevates the Gibbs free energy and moves the $G=0$ crossing, reflecting how Lorentz-violation-induced rescaling modifies the global thermodynamic preference. Panel~(b) indicates that increasing $\eta$ tends to lower $G$, consistent with the monopole-induced deficit reducing the effective gravitational contribution. Panel~(c) demonstrates that increasing $\gamma$ suppresses the electromagnetic term $3e^{-\gamma}Q^2/[(1-\ell)^2 r_h^2]$, which primarily affects the small-$r_h$ regime and hence changes the onset of global stability.

Finally, for the KR-AdS black hole spacetime endowed with a global monopole, we evaluate the specific heat capacity at constant pressure, $C_P$, which serves as a key diagnostic of the local thermodynamic stability of the black hole. In the framework of extended phase space thermodynamics, the specific heat at constant pressure is defined as
\begin{equation}
C_P= \left(\frac{\partial M}{\partial T}\right)_P= -2\pi r_h^2
\frac{\left[\frac{1-k\eta^2}{1-\ell}-\frac{e^{-\gamma}Q^2}{(1-\ell)^2 r_h^2}
+8\pi P r_h^2\right]}{\left[\frac{1-k\eta^2}{1-\ell}-\frac{3e^{-\gamma}Q^2}{(1-\ell)^2 r_h^2}
-8\pi P r_h^2\right]}.
\label{dd7}
\end{equation}
A positive value of $C_P$ corresponds to a locally stable thermodynamic phase, whereas a negative value signals instability. Moreover, the divergence of $C_P$ marks the occurrence of a second-order phase transition separating distinct black hole phases.

\begin{figure}[htb!]
\centering
\includegraphics[width=\textwidth]{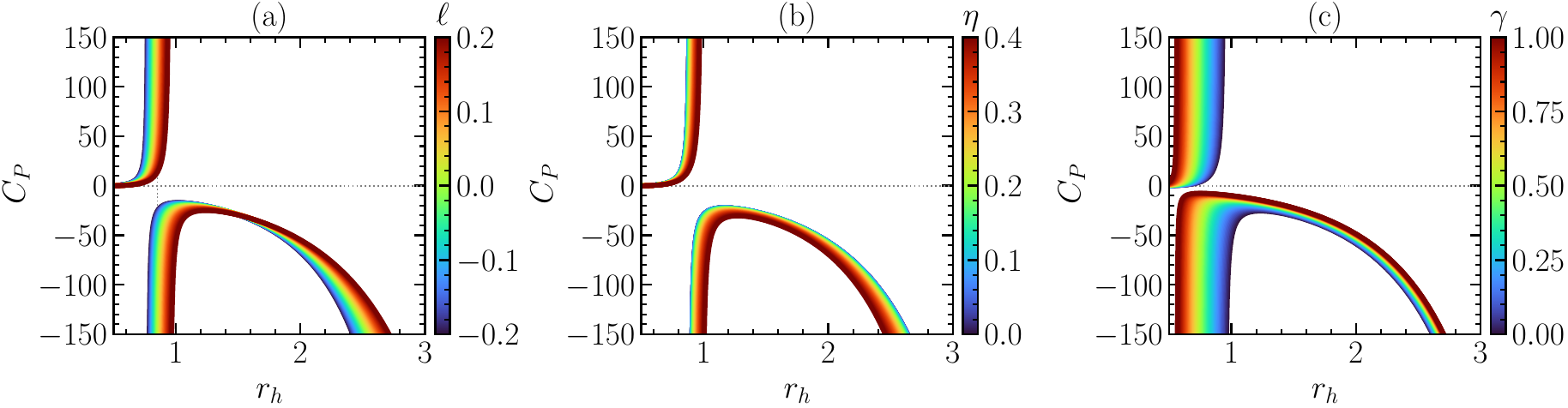}
\caption{Specific heat at constant pressure $C_{P}$ as a function of the event-horizon radius $r_h$ [Eq.~(\ref{dd7})], for $Q=0.5$, $k=1$, and $P=0.003$. In each panel, one parameter is varied continuously and encoded by the color gradient (see colorbar), while the remaining parameters are fixed. Panel~(a): $\ell$ varied continuously (with $\eta=0.1$ and $\gamma=0.1$ fixed). Panel~(b): $\eta$ varied continuously (with $\ell=0.1$ and $\gamma=0.1$ fixed). Panel~(c): $\gamma$ varied continuously (with $\ell=0.1$ and $\eta=0.1$ fixed). Divergences of $C_{P}$ (at fixed $P$) indicate second-order phase transitions separating small, intermediate, and large black-hole branches.}
\label{fig:specific_heat}
\end{figure}

The specific heat at constant pressure is shown in Fig.~\ref{fig:specific_heat}. In each panel, $C_P(r_h)$ exhibits the characteristic divergence structure of charged-AdS black holes, with divergences determined by the vanishing of the denominator in Eq.~(\ref{dd7}). The continuous color mapping provides a compact visualization of how the divergence locations migrate as the model parameters vary smoothly. In panel~(a), varying $\ell$ shifts the divergence points and modifies the width of the unstable intermediate branch. In panel~(b), varying $\eta$ produces a similar qualitative effect through the combination $(1-k\eta^2)$, typically with a milder dependence across the explored interval. In panel~(c), varying $\gamma$ mainly reshapes the inner (small-$r_h$) divergence because $\gamma$ directly suppresses the charge contribution, while the outer divergence is comparatively less sensitive.

\begin{figure}[htb!]
\centering
\includegraphics[width=\textwidth]{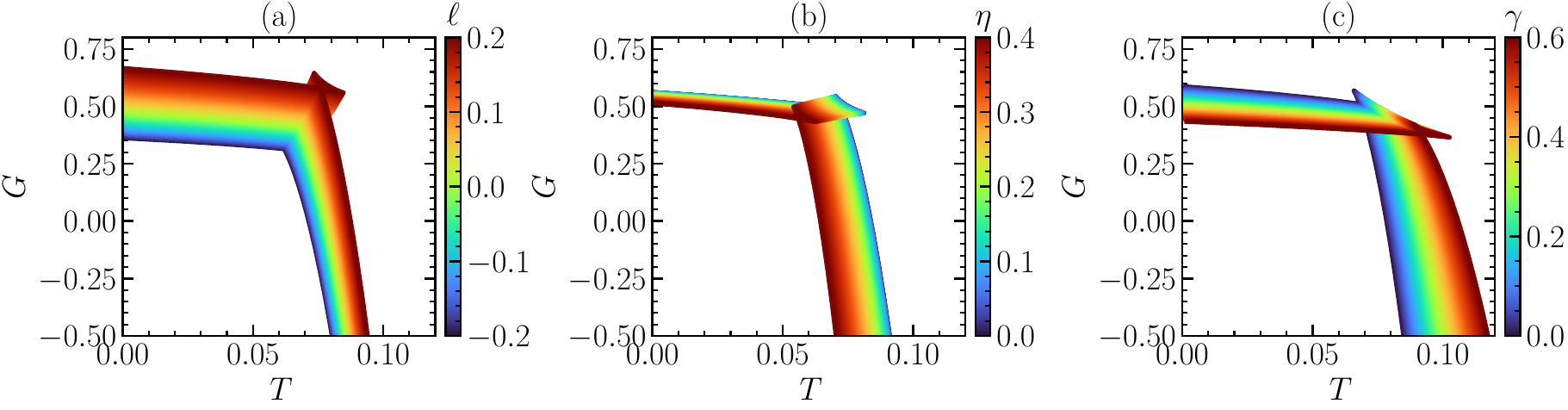}
\caption{Gibbs free energy $G$ versus Hawking temperature $T$ (parametric in $r_h$) at pressure $P=0.5\,P_c$, where $P_c$ is computed for each parameter set using the corresponding critical condition. The swallowtail behavior signals a first-order phase transition. In each panel, one parameter is varied continuously and encoded by the color gradient (see colorbar), while the remaining parameters are fixed. Panel~(a): $\ell$ varied continuously (with $\eta=0.1$ and $\gamma=0.1$ fixed). Panel~(b): $\eta$ varied continuously (with $\ell=0.1$ and $\gamma=0.1$ fixed). Panel~(c): $\gamma$ varied continuously (with $\ell=0.1$ and $\eta=0.1$ fixed).}
\label{fig:gibbs_T}
\end{figure}

The global phase structure is further elucidated through the $G$--$T$ diagram presented in Fig.~\ref{fig:gibbs_T}, where the Gibbs free energy is plotted as a parametric function of the Hawking temperature, with $r_h$ serving as the parameter and the pressure fixed below the critical value ($P < P_c$). The characteristic swallowtail structure, hallmark of a first-order phase transition between small and large black holes, is clearly visible in all panels. In the updated visualization, the relevant model parameter is varied continuously (as indicated by the panel-specific colorbar), which makes the deformation of the swallowtail envelope and the transition point a smooth function of the parameter. Panel~(a) shows that increasing $\ell$ shifts the swallowtail to higher temperatures and changes the overall Gibbs scale. Panel~(b) shows that increasing $\eta$ shifts the swallowtail through the monopole-induced geometric factor. Panel~(c) shows that varying $\gamma$ reshapes the swallowtail by suppressing the electromagnetic sector, thereby altering the multivalued region and the first-order transition temperature. At $P = P_c$, the swallowtail degenerates into a cusp (second-order critical point), and for $P > P_c$ the $G(T)$ curve becomes single-valued, indicating a smooth crossover with no phase transition.

We now focus on the first law of thermodynamics for a charged AdS black hole and examine the corresponding Smarr relation in terms of the thermodynamic variables $S$, $P$, and $Q$. The black hole mass $M$, expressed as a function of the entropy, is given by
\begin{equation}
 M=\frac{1}{2}\sqrt{S/\pi}\,\left[\frac{1-k \eta^2}{1 - \ell}+e^{-\gamma}\frac{\pi Q^2_{\rm eff}}{S}+\frac{8P}{3} S\right].\label{dd8}
\end{equation}
The first law of black hole thermodynamics in extended phase space is written as:
\begin{equation}
    dM=T\,dS+V\,dP+\Phi\,dQ_{\rm eff},\label{dd9}
\end{equation}
where $T$ is the temperature, $V$ is the thermodynamic volume and $\Phi$ is the electric potential. These are given by 
\begin{align}
    &V=\left(\frac{\partial M}{\partial P}\right)_{S, Q}=\frac{4}{3}\sqrt{\frac{S^3}{\pi}}=\frac{4\pi}{3} r^3_h,\label{dd9b}\\
    &\Phi=\left(\frac{\partial M}{\partial Q_{\rm eff}}\right)_{S, P}=e^{-\gamma}\frac{\pi Q_{\rm eff}}{\sqrt{\pi S}}=e^{-\gamma}\frac{Q_{\rm eff}}{r_h}.\label{dd10}
\end{align}
Applying Euler's theorem for homogeneous functions, one can show that the Smarr relation given below satisfies in our case
\begin{equation}
M = 2\,T\,S - 2\,P\,V + \Phi\,Q_{\rm eff}.
\end{equation} 
The above Smarr formula demonstrates that the black hole mass should be interpreted as the enthalpy of the spacetime rather than the internal energy when the cosmological constant is treated as a dynamical pressure. Moreover, the appearance of the effective charge $Q_{\rm eff}$ indicates that the electromagnetic sector modifies the global energy balance without altering the overall scaling structure of the Smarr relation.

\section{$P$-$v$ Criticality}

In black hole thermodynamics, a critical point marks the termination of a first-order phase transition, such as the transition between small and large black holes. This phenomenon is closely analogous to the liquid-gas critical point encountered in ordinary thermodynamic systems \cite{Kubiznak2012, Mann2014, Kastor2009}. At the critical point, fundamental thermodynamic quantities, including the black hole temperature, pressure, and volume, satisfy specific conditions that unveil universal features reminiscent of conventional fluids. The transition between small and large black holes can be effectively captured using a Van der Waals-type equation of state, where the black hole's pressure, volume, and temperature mimic the behavior of a real fluid, exhibiting characteristic critical phenomena such as the inflection point in the $P$-$v$ curve at the critical temperature. This analogy provides a powerful tool to study black hole thermodynamics and to uncover universal scaling relations \cite{Kubiznak2012, Mann2014}.

By introducing the specific volume $v = 2 r_h$ and expressing the pressure $P$ in terms of the Hawking temperature $T$, Eq.~(\ref{dd4}) can be written as
\begin{align}
P(T, v)= \frac{T}{v} - \frac{\mathcal{C}}{v^2}+ \frac{\mathcal{D}}{v^4},\label{ee1}
\end{align}
where
\begin{equation}
\mathcal{C}=\frac{1-k \eta^2}{2\pi(1-\ell)},\qquad 
\mathcal{D}=\frac{2 e^{-\gamma}\,Q^2}{\pi (1-\ell)^2}.\label{ee2}
\end{equation}

The location of the critical point is determined from the standard inflection point conditions of the $P$-$v$ diagram:
\begin{equation}
\left(\frac{\partial P}{\partial v}\right)_{T}=0, 
\qquad
\left(\frac{\partial^{2} P}{\partial v^{2}}\right)_{T}=0.
\label{ee3}
\end{equation}
Using (\ref{ee1}) into these condition and after simplification yields the critical volume, temperature and pressure as follows:
\begin{align}
v_c &= \sqrt{\frac{6 \mathcal{D}}{\mathcal{C}}}=\sqrt{\frac{24\, e^{-\gamma} Q^2}{(1-\ell)(1-k \eta^2)}},\nonumber\\[2mm]
T_c &= \frac{4 \mathcal{C}^{3/2}}{3 \sqrt{6 \mathcal{D}}}=\frac{(1-k \eta^2)^{3/2}}{3\pi e^{-\gamma/2}  Q \sqrt{6(1-\ell)}},\nonumber\\[2mm]
P_c &= \frac{\mathcal{C}^2}{12 \mathcal{D}}=\frac{(1-k \eta^2)^2}{96 \pi\, e^{-\gamma} Q^2}.\label{ee4}
\end{align}
The critical ratio is given by
\begin{equation}
    \rho_c=\frac{P_c v_c}{T_c}=3/8,\label{ee5}
\end{equation}
which is similar to the Van der Waals ratio for ideal fluid.

In the limiting case where the global monopole is absent ($\eta=0$), the KR field vanishes ($\ell=0$), and the ModMax parameter is set to zero ($\gamma=0$), the critical points reduce to the well-known values for the standard Reissner–Nordström–AdS black hole:
\begin{align}
v_c &= 2\sqrt{6}\, Q, \nonumber\\[1mm]
T_c &= \frac{1}{3 \sqrt{6} \pi Q}, \nonumber\\[1mm]
P_c &= \frac{1}{96 \pi\, Q^2}. \label{ee6}
\end{align}
These expressions match the results originally reported in \cite{Kubiznak2012}, confirming that in the absence of additional fields or parameters, the KR–AdS black hole reduces to the familiar RN-AdS case.

\begin{figure}[htb!]
\centering
\includegraphics[width=\textwidth]{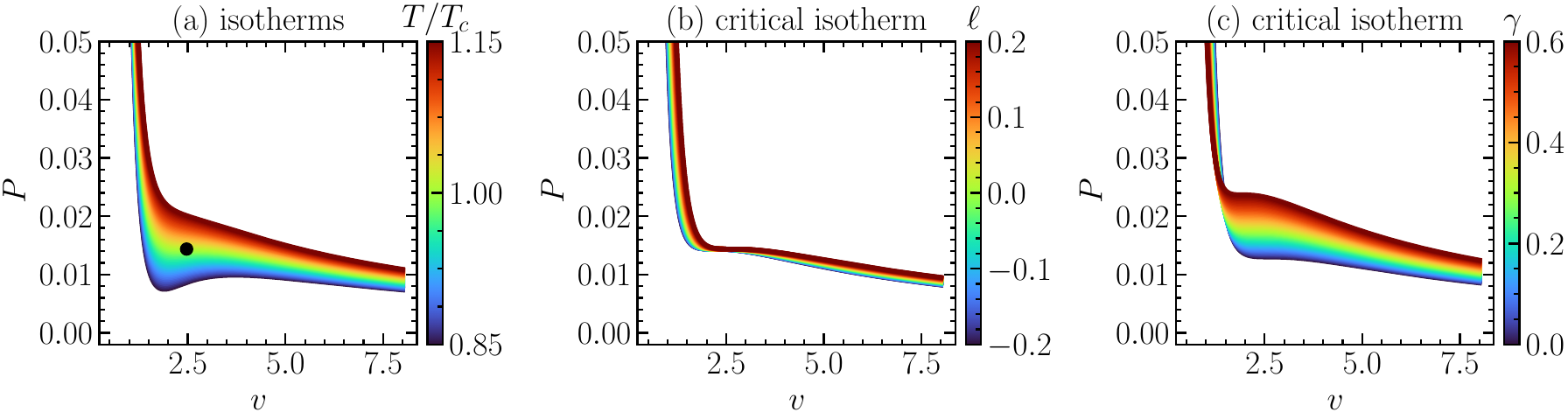}
\caption{$P$--$v$ isotherms obtained from the equation of state [Eq.~(\ref{ee1})]. Panel~(a): isotherms at $T/T_c=0.85$, $1.0$, and $1.15$ for $\ell=0.1$, $\eta=0.1$, $\gamma=0.1$, $Q=0.5$, and $k=1$; the black dot marks the critical point $(v_c,P_c)$ computed from Eq.~(\ref{ee4}). Panel~(b): critical isotherms at $T=T_c$ with $\ell$ varied continuously (color-coded; $\eta=0.1$, $\gamma=0.1$, $Q=0.5$, $k=1$ fixed). Panel~(c): critical isotherms at $T=T_c$ with $\gamma$ varied continuously (color-coded; $\ell=0.1$, $\eta=0.1$, $Q=0.5$, $k=1$ fixed).}
\label{fig:Pv_criticality}
\end{figure}

The $P$-$v$ isotherms and the critical behavior are illustrated in Fig.~\ref{fig:Pv_criticality}. Panel~(a) displays isotherms at three representative temperatures relative to the critical temperature $T_c$: below ($T/T_c=0.85$), at ($T/T_c=1.0$), and above ($T/T_c=1.15$). Below the critical temperature, the isotherm exhibits a Van der Waals-like oscillatory behavior with a local maximum and minimum, signaling the coexistence of small and large black hole phases separated by an unstable intermediate branch. At $T=T_c$, the inflection point (marked by the black dot) emerges, and for $T>T_c$ the isotherm becomes monotonically decreasing, indicating a supercritical regime with no phase transition. Panel~(b) and panel~(c) show how the \emph{critical isotherm} at $T=T_c$ is deformed when the geometric parameters are varied continuously: the color gradient encodes the varying parameter and allows one to follow the smooth drift of the critical point and the overall equation-of-state profile across the chosen interval.

\begin{figure}[htb!]
\centering
\includegraphics[width=\textwidth]{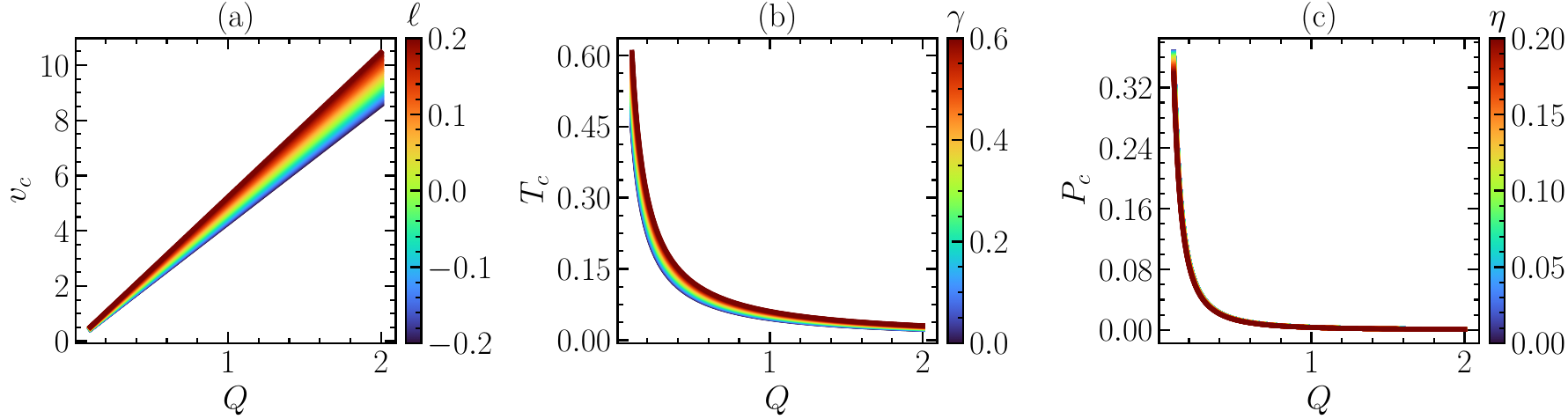}
\caption{Critical quantities as functions of the electric charge $Q$ computed from Eq.~(\ref{ee4}), for $\eta=0.1$ and $k=1$. Panel~(a): critical specific volume $v_c(Q)$. Panel~(b): critical temperature $T_c(Q)$. Panel~(c): critical pressure $P_c(Q)$. Curves correspond to the parameter sets indicated in the figure legend.}
\label{fig:critical_quantities}
\end{figure}

The dependence of the critical quantities $v_c$, $T_c$, and $P_c$ on the electric charge $Q$ is presented in Fig.~\ref{fig:critical_quantities}. In the updated plots, the different parameter choices are represented in a consistent way (as shown in the figure), keeping the focus on the functional dependence of the critical quantities on $Q$. Panel~(a) shows that the critical specific volume $v_c$ grows with $Q$ as expected from $v_c \propto Q/\sqrt{(1-\ell)(1-k\eta^2)}$. Panel~(b) demonstrates that the critical temperature $T_c$ decreases with $Q$ as $T_c\propto 1/Q$. Panel~(c) shows that the critical pressure $P_c$ decreases as $Q^{-2}$ and is strongly affected by $\gamma$ through $P_c\propto 1/[e^{-\gamma}Q^2]$, so that increasing $\gamma$ enhances $P_c$ at fixed $Q$. These results confirm that the universal critical ratio $\rho_c=3/8$ is preserved, while the critical scales are tunable by the spacetime and electromagnetic parameters.

\section{Joule-Thomson Expansion: Inversion Temperature}

The Joule-Thomson (JT) expansion is a thermodynamic process in which a system undergoes adiabatic expansion through a porous plug or throttle, leading to a change in temperature at constant enthalpy. In the context of black hole thermodynamics, the mass of the black hole is identified with the enthalpy $M$ in extended phase space, allowing the study of JT expansion for black holes \cite{Okcu2017,Mo2018}. The JT coefficient is defined as
\begin{equation}
\mu_{\rm JT} = \left(\frac{\partial T}{\partial P}\right)_M,\label{ff1}
\end{equation}
and its sign determines whether the black hole cools ($\mu_{\rm JT} > 0$) or heats ($\mu_{\rm JT} < 0$) during expansion. The \emph{inversion temperature} $T_i$ corresponds to the condition $\mu_{\rm JT}=0$, separating the cooling and heating regions. For charged AdS black holes, $T_i$ depends on the black hole charge, horizon radius, and other parameters, and the inversion curve in the $T$--$P$ plane provides insights into the thermodynamic behavior analogous to classical fluids \cite{YasirJaved2024,Ahmed2025}. Studying the JT expansion and inversion temperature in modified theories of gravity, such as those including global monopoles, KR fields, or nonlinear electrodynamics, reveals rich phenomena including multiple inversion points, modified cooling-heating regimes, and shifts in critical behavior compared to the standard RN-AdS case.

Recall the expression for the Joule Thomson coefficient
 \begin{equation}
  \mu_{JT}=\frac{1}{C_{P}}\left[T \left(\frac{\partial V}{\partial T}\right)_P-V\right].\label{ff2}   
 \end{equation} 
For zero JT-coefficient ($\mu_J=0$), one can find inversion temperature as
\begin{equation}
    T_i=V \left(\frac{\partial T}{\partial V}\right)_{P=P_i}.\label{ff3} 
\end{equation}
In our case at hand, we find
\begin{equation}
T_i=\frac{1}{12 \pi} \left[ - \frac{1-k\eta^2}{(1-\ell) r_h} + \frac{3 e^{-\gamma} Q^2}{(1-\ell)^2 r_h^3} + 8 \pi P_i r_h \right].\label{ff4}
\end{equation}

Recall the Hawking temperature given in Eq. (\ref{dd4}) at $P=P_i$ given by
\begin{equation}
T_i=\frac{1}{4\pi r_h}\left[\frac{1-k \eta^2}{1-\ell}-e^{-\gamma}\,\frac{Q^2}{(1-\ell)^2 r^2_h}+8\pi P_i\,r^2_h\right].\label{ff5}
\end{equation}

Subtracting Eq. (\ref{ff4}) from (\ref{ff5}) we find (setting $x=r^2_h$)
\begin{equation}
\frac{1-k \eta^2}{1-\ell} x - \frac{3 e^{-\gamma} Q^2}{2(1-\ell)^2} + 4 \pi P_i x^2=0.\label{ff6}
\end{equation}
The real and positive root of this quadratic equation is given by
\begin{equation}
r_h=\frac{1}{2\sqrt{2 \pi\,P_i}}\sqrt{\frac{\sqrt{(1-\kappa \eta^2)+24 \pi P_i e^{-\gamma} Q^2}}{1-\ell}-\frac{(1-\kappa \eta^2)}{1-\ell}}.\label{ff7}
\end{equation}

Substituting $r_h$ into the Eq. (\ref{ff5}) results
\begin{align}
T_i=\frac{\sqrt{P_i}}{\sqrt{2 \pi}}\,\frac{\Bigg[\frac{1-k \eta^2}{1-\ell}-e^{-\gamma}\,\frac{ 8\pi P_i Q^2}{(1-\ell)^2}\,\frac{1}{\frac{\sqrt{(1-\kappa \eta^2)+24 \pi P_i e^{-\gamma} Q^2}}{1-\ell}-\frac{(1-\kappa \eta^2)}{1-\ell}}
+\frac{\sqrt{(1-\kappa \eta^2)+24 \pi P_i e^{-\gamma} Q^2}}{1-\ell}-\frac{(1-\kappa \eta^2)}{1-\ell}\Bigg]}{\sqrt{\frac{\sqrt{(1-\kappa \eta^2)+24 \pi P_i e^{-\gamma} Q^2}}{1-\ell}-\frac{(1-\kappa \eta^2)}{1-\ell}}}.\label{ff8}
\end{align}

At zero inversion pressure $P_i=0$, from Eq (\ref{ff6}) one finds
\begin{equation}
r^{\rm min}_h=\sqrt{\frac{3 e^{-\gamma} }{2 (1-\kappa \eta^2)(1-\ell)}}\,Q.\label{ff9}
\end{equation}
Therefore, minimum inversion temperature from (\ref{ff5}) is as follows:
\begin{equation}
T^{\rm min}_i=\frac{e^{\gamma/2}\,(1-\kappa \eta^2)^{3/2}}{6 \pi\sqrt{6 (1-\ell)}\,Q}.\label{ff10}
\end{equation}

\begin{figure}[htb!]
\centering
\includegraphics[width=\textwidth]{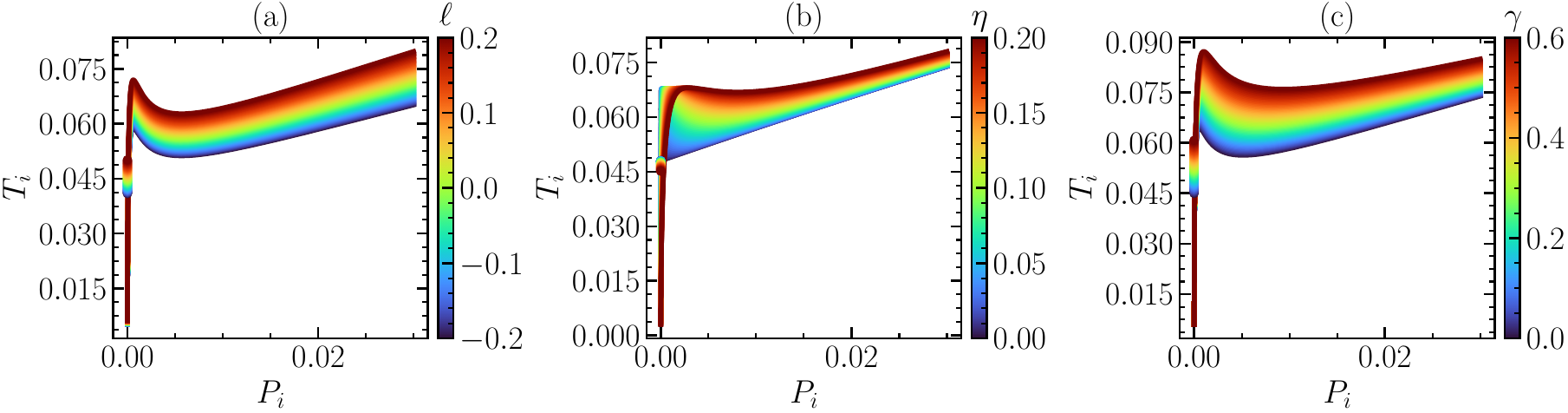}
\caption{Joule--Thomson inversion curves $T_i(P_i)$ obtained from Eqs.~(\ref{ff5})--(\ref{ff8}) for $Q=0.5$ and $k=1$. In each panel, one parameter is varied continuously and encoded by the color gradient (see colorbar), while the remaining parameters are fixed. Panel~(a): $\ell$ varied continuously (with $\eta=0.1$ and $\gamma=0.1$ fixed). Panel~(b): $\eta$ varied continuously (with $\ell=0.1$ and $\gamma=0.1$ fixed). Panel~(c): $\gamma$ varied continuously (with $\ell=0.1$ and $\eta=0.1$ fixed). The colored dots at $P_i=0$ mark the minimum inversion temperature $T_i^{\rm min}$ given by Eq.~(\ref{ff10}). For a given pressure, the region above (below) the inversion curve corresponds to cooling (heating), i.e., $\mu_{\rm JT}>0$ ($\mu_{\rm JT}<0$).}
\label{fig:JT_inversion}
\end{figure}

The Joule--Thomson inversion curves in the $T_i$--$P_i$ plane are presented in Fig.~\ref{fig:JT_inversion}. Each curve is obtained by substituting the horizon radius from Eq.~(\ref{ff7}) into the Hawking temperature Eq.~(\ref{ff5}), generating the locus of points where $\mu_{\rm JT}=0$. Above the inversion curve, the JT coefficient is positive ($\mu_{\rm JT}>0$) and the black hole cools during isenthalpic expansion; below it, the black hole heats ($\mu_{\rm JT}<0$). In the updated figure, the varied parameter is encoded continuously by the color gradient, and the minimum inversion temperature $T_i^{\rm min}$ at $P_i=0$ [Eq.~(\ref{ff10})] is indicated by the colored markers on the vertical axis. This representation emphasizes the smooth upward/downward drift of both the inversion curve and $T_i^{\rm min}$ as each parameter changes within the adopted interval.

\section{Sparsity of Hawking Radiation}

\noindent In this section, we evaluate the sparsity of the Hawking radiation emitted by our black hole solution. Although a black hole radiates thermally with a temperature determined by its surface gravity, mimicking a classical black body the Hawking flux is discrete in time: emission occurs in well-separated quanta rather than as a continuous flow. Sparsity quantifies this intermittent character by comparing the square of the thermal wavelength \(\lambda_T=2\pi/T_H\) to the effective emitting area \(\mathcal{A}_{\rm eff}\) and is conventionally expressed as \begin{equation}
\label{defSpars}
    \psi =\frac{\mathcal{C}}{\Tilde{g} }\left(\frac{\lambda_t^2}{\mathcal{A}_{\rm eff}}\right),
\end{equation}
where \(\mathcal{C}\) is a dimensionless constant and \(\tilde g\) the spin degeneracy of the emitted quanta \cite{HawkingPage,FG2016,YS2025}. For a Schwarzschild black hole \(T_H=1/(4\pi r_h)\), so \(\lambda_T=2\pi/T_H=8\pi^2 r_h\), while \(\mathcal{A}_{\rm eff}=(27/4)\mathcal{A}_{\rm BH}=27\pi r_h^2\). Substituting these expressions into \eqref{defSpars} yields the closed form
\begin{equation}
\psi_{\rm Sch}=\frac{\mathcal{C}}{\tilde g}\,\frac{\lambda_T^2}{\mathcal{A}_{\rm eff}}
=\frac{\mathcal{C}}{\tilde g}\,\frac{64\pi^3}{27}\simeq\frac{\mathcal{C}}{\tilde g}\times 73.49.
\end{equation}
Hence, for the canonical choice \(C/\tilde g=1\) one obtains \(\psi_{\rm Sch}\approx73.5\) (for photon emission with \(\tilde g=2\) this value is halved), demonstrating that Hawking radiation is extremely sparse (\(\psi\gg1\)) in stark contrast to ordinary black-body emission where \(\psi\ll1\).  In the realm of our black hole solution the modified surface gravity and horizon geometry enter both \(T\) and \(\mathcal{A}_{\rm eff}\), so one naturally expects the sparsity parameter \(\psi\) to exhibit nontrivial dependence on \((k\,\eta^2,\ell)\), with direct consequences for the discreteness and observational character of the evaporation process.\\ Substituting the temperature \eqref{dd4} into the sparsity definition \eqref{defSpars} yields the closed form
\begin{equation}\label{sspp}
\psi(r_h)=\frac{64\, \pi ^3 \,e^{2\gamma}(1-\ell)^4\,r_h^4}{27 \left(Q^2 +e^\gamma (1-\ell) r_h^2 \left(k\eta^2-1+\Lambda  r_h^2\right)\right)^2}.\end{equation}

From the above expression, it is evident that the sparsity behavior of the black hole depends on all the geometric parameters $(Q,\, \ell,\, \eta,\, \Lambda,\, \gamma)$ characterizing the spacetime. In the limit $Q=0,\,\ell=0,\,\eta=0,\,\Lambda=0$, one recovers the Schwarzschild black hole result $\psi \to \psi_{\rm Sch}$.
\begin{figure*}[tbhp]
\centering
\includegraphics[width=\textwidth]{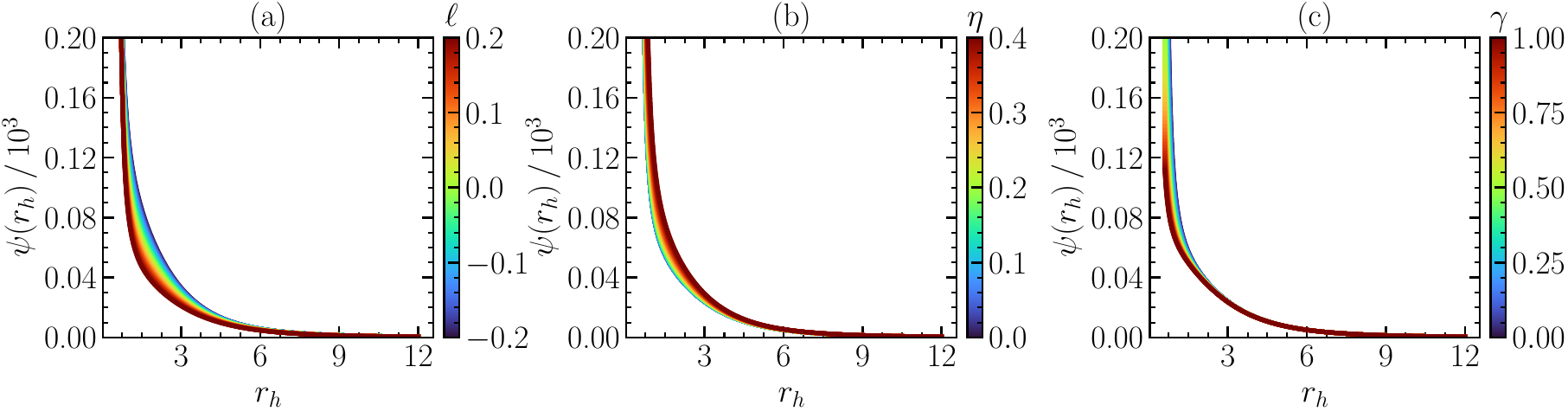}
\caption{Sparsity parameter $\psi(r_h)$ [Eq.~(\ref{sspp})] as a function of the horizon radius $r_h$, displayed in units of $10^{3}$ (vertical axis shows $\psi(r_h)/10^{3}$). The three panels illustrate the sensitivity of $\psi(r_h)$ to each model parameter by varying one parameter continuously while keeping the others fixed at the reference values $Q=0.5$, $k=1$, $P=0.003$, and $(\ell,\eta,\gamma)=(0.1,0.1,0.1)$. Panel~(a): $\ell\in[-0.2,0.2]$ (color scale). Panel~(b): $\eta\in[0,0.4]$ (color scale). Panel~(c): $\gamma\in[0,1]$ (color scale). In all panels, the color gradient encodes the continuously varying parameter, as indicated by the corresponding vertical colorbar.}
    \label{fig:psi_eq42}
\end{figure*}
Figure~\ref{fig:psi_eq42} shows the radial dependence of the sparsity parameter $\psi(r_h)$ defined in Eq.~(\ref{sspp}). 
For the parameter ranges considered, $\psi(r_h)$ is a positive, rapidly decaying function of the horizon radius: it attains its largest values for small $r_h$ and decreases monotonically as $r_h$ increases, approaching a near-zero baseline for $r_h\gtrsim\mathcal{O}(5\text{--}10)$. 
The continuous color encoding allows one to track the systematic impact of each parameter without cluttering the panels with multiple discrete legends. 
In Fig.~\ref{fig:psi_eq42}(a), varying $\ell$ within $[-0.2,0.2]$ produces the most visible spread at small $r_h$, indicating that the Lorentz-violating deformation can appreciably modulate $\psi(r_h)$ close to the horizon, while the curves converge for large radii. 
Figure~\ref{fig:psi_eq42}(b) exhibits a milder but still noticeable dependence on the global monopole parameter $\eta$ in the same near-horizon region, whereas the asymptotic decay at larger $r_h$ remains essentially unchanged across the interval $\eta\in[0,0.4]$. 
Finally, Fig.~\ref{fig:psi_eq42}(c) shows that the ModMax parameter $\gamma$ produces only a modest variation in the profile over $\gamma\in[0,1]$, with the dominant qualitative behavior, a steep near-horizon drop followed by a long tail toward zero, preserved throughout. 
Overall, Fig.~\ref{fig:psi_eq42} indicates that the sparsity diagnostic is primarily controlled by the near-horizon regime, where parameter changes are most effective, while the large-$r_h$ behavior is comparatively robust within the explored domain.

\section{Emission energy}

Quantum effects in curved spacetime imply that black holes are not perfectly black: a thermal flux is emitted from the near-horizon region with temperature fixed by the surface gravity. In semiclassical language, this phenomenon may be interpreted through tunneling/particle production near the event horizon, which leads to a gradual decrease of the black-hole mass as energy is carried away to infinity \cite{Javed2019}.

In the geometric-optics (high-frequency) regime, the absorption cross section oscillates around a constant limiting value $\sigma_{\rm lim}$. Motivated by the fact that the capture of high-energy quanta is governed by null geodesics, one may approximate this limiting cross section by a horizon-scale area, $\sigma_{\rm lim}\approx \pi r_h^2$ \cite{Ditta2022}, where $r_h$ is the event-horizon radius. Within this approximation, the spectral energy emission rate takes the standard blackbody-like form \cite{Mustafa2025}
\begin{equation}
\frac{d^{2}\varepsilon }{d\omega\,dt}
=\frac{2\pi ^{2}\sigma _{\rm lim}}{e^{\omega/T}-1}\,\omega ^{3},
\label{eee1}
\end{equation}
where $\omega$ is the emitted frequency and $T$ is the Hawking temperature.

For the ModMax--AdS black hole with global monopole in KR-gravity, the temperature is given in Eq.~(\ref{dd4}). Substituting Eq.~(\ref{dd4}) into Eq.~(\ref{eee1}) and using $\sigma_{\rm lim}\approx \pi r_h^2$, we obtain the explicit parameter-dependent emission rate
\begin{equation}
\frac{d^{2}\varepsilon }{d\omega\,dt}
=2\pi ^{3}\,r_h^2\,\omega^3\,
\left[
\exp\left\{
4 \pi r_h\,\omega\,
\left(
\frac{1-k \eta^2}{1-\ell}
-e^{-\gamma}\,\frac{Q^2}{(1-\ell)^2 r^2_h}
+8\pi P\,r^2_h
\right)^{-1}
\right\}-1
\right]^{-1}.
\label{eee2}
\end{equation}
Equation~(\ref{eee2}) shows explicitly how the geometric parameters and the nonlinear electromagnetic sector influence the Hawking output through the temperature. In particular, the Lorentz-violating parameter $\ell$ and the global monopole parameter $\eta$ enter through the combination $(1-k\eta^2)/(1-\ell)$, whereas the ModMax parameter $\gamma$ exponentially suppresses the effective charge contribution via $e^{-\gamma}Q^2$. Finally, the pressure $P$ controls the large-$r_h$ behavior through the AdS term. Therefore, modifications in $(\ell,\eta,\gamma)$ can shift the spectral peak and change the overall intensity of the emission, especially in the small-$r_h$ regime where the charge sector competes with the geometric term in Eq.~(\ref{dd4}).

\begin{figure*}[htb!]
\centering
\includegraphics[width=\textwidth]{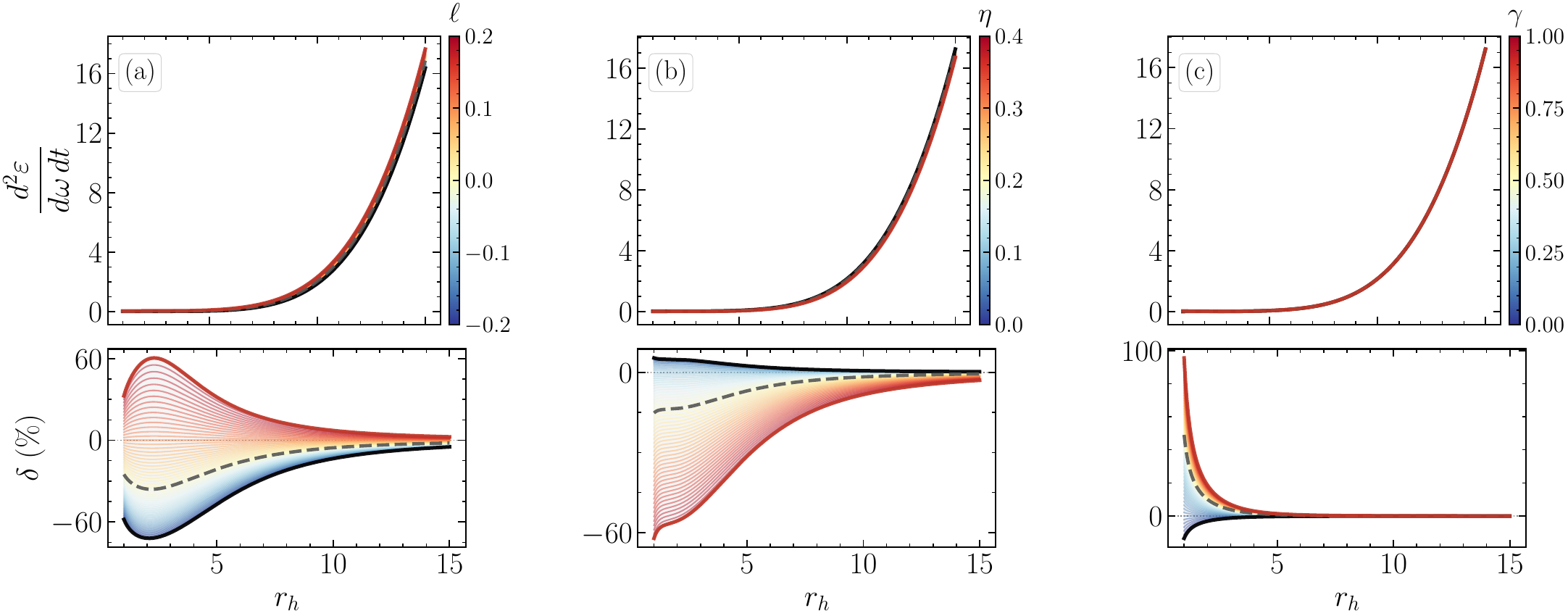}
\caption{Spectral energy emission rate $d^2\varepsilon/(d\omega\,dt)$ [Eq.~(\ref{eee2})] as a function of the event-horizon radius $r_h$ at fixed frequency $\omega=0.3$, for $Q=0.5$, $k=1$, and $P=0.003$. \textbf{Top row:} emission rate. \textbf{Bottom row:} relative percentage deviation $\delta(\%)$ with respect to the reference parameter set $(\ell_0,\eta_0,\gamma_0)=(0.1,0.1,0.1)$. In each column, one parameter is varied continuously and encoded by the color gradient (see colorbar), while the remaining parameters are fixed at their reference values. Three anchor curves (minimum, reference, and maximum of the swept parameter) are highlighted for guidance. Panel~(a): $\ell\in[-0.2,\,0.2]$ (with $\eta=0.1$ and $\gamma=0.1$ fixed). Panel~(b): $\eta\in[0,\,0.4]$ (with $\ell=0.1$ and $\gamma=0.1$ fixed). Panel~(c): $\gamma\in[0,\,1]$ (with $\ell=0.1$ and $\eta=0.1$ fixed). The shaded regions in the residual subpanels indicate the full parameter envelope.}
\label{fig:emission}
\end{figure*}

In Fig.~\ref{fig:emission}, we display the spectral energy emission rate $d^2\varepsilon/(d\omega\,dt)$ as a function of the event-horizon radius $r_h$ at a fixed representative frequency $\omega=0.3$, using the full expression in Eq.~(\ref{eee2}). The top row of panels presents the emission rate itself, while the bottom row shows the relative percentage deviation $\delta(\%)=[y(\text{param})-y(\text{ref})]/y(\text{ref})\times 100$ with respect to the reference parameter set $(\ell_0,\eta_0,\gamma_0)=(0.1,0.1,0.1)$. In each column, one parameter is varied continuously over a prescribed interval and encoded by the color gradient (see the vertical colorbar), while the remaining parameters are held fixed at their reference values; additionally, three anchor curves corresponding to the minimum, reference, and maximum values of the swept parameter are highlighted. Across all panels, the emission rate is a monotonically increasing function of $r_h$, reflecting the combined growth of the effective emitting area $\sigma_{\rm lim}=\pi r_h^2$ and the Hawking temperature in the pressure-dominated regime at large $r_h$.

Figure~\ref{fig:emission}(a) illustrates the effect of the Lorentz-violating parameter $\ell\in[-0.2,\,0.2]$: the resulting spread in the emission rate is approximately symmetric about the reference curve, reaching relative deviations of order $\pm\,50\%$ at small $r_h\lesssim 3$ and converging to $|\delta|\lesssim 5\%$ for $r_h\gtrsim 10$. This behavior reflects the fact that $\ell$ enters the temperature primarily through the prefactor $(1-\ell)^{-1}$, which produces a multiplicative rescaling whose effect is amplified where the emission rate itself is small and the Boltzmann suppression factor $e^{\omega/T}$ is most sensitive to small changes in $T$. In the residual subpanel, the envelope formed by the color-coded curves clearly demonstrates this convergence at large radii.

Figure~\ref{fig:emission}(b) shows the influence of the global monopole parameter $\eta\in[0,\,0.4]$: the deviation is predominantly negative and reaches $\delta\approx -60\%$ at large $r_h$. Increasing $\eta$ reduces the effective geometric factor $(1-k\eta^2)/(1-\ell)$ in Eq.~(\ref{dd4}), thereby lowering the Hawking temperature and, consequently, suppressing the emission rate. Unlike the case of $\ell$, the deviation does not converge at large radii but instead persists because the factor $(1-k\eta^2)$ enters linearly in the dominant large-$r_h$ contribution to $T$.

figure~\ref{fig:emission}(c) depicts the role of the ModMax parameter $\gamma\in[0,\,1]$: the deviation is positive and can exceed $90\%$ at small $r_h$, decaying rapidly to $\lesssim 5\%$ for $r_h\gtrsim 7$. The exponential suppression factor $e^{-\gamma}$ reduces the effective charge contribution $e^{-\gamma}Q^2/[(1-\ell)^2 r_h^2]$ in Eq.~(\ref{dd4}), which is a negative term in the temperature bracket. Increasing $\gamma$ therefore raises $T$ and enhances the emission rate. Since the charge term scales as $r_h^{-2}$, its influence (and hence the sensitivity to $\gamma$) is concentrated in the small-$r_h$ regime, consistent with the rapid decay of $\delta$ observed in the residual subpanel.

Overall, Fig.~\ref{fig:emission} demonstrates that the emission energy rate inherits a rich parameter dependence from the Hawking temperature structure: $\ell$ produces a symmetric, convergent perturbation; $\eta$ induces a persistent suppression that grows with $r_h$; and $\gamma$ generates a localized enhancement confined to the near-horizon regime. These features have direct implications for the evaporation dynamics of the black hole, as the parameter-dependent modulation of the spectral output alters both the total radiated power and the timescale over which the black hole loses mass.

\section{Thermal fluctuations}

We now examine the impact of thermal fluctuations on the entropy of the black hole. In the Euclidean quantum gravity approach, one computes the partition function by Wick rotating the time coordinate and expanding the action around the saddle-point geometry \cite{Pourhassan2015}. In this framework, thermal/quantum fluctuations around equilibrium generate corrections to the leading semiclassical entropy. Such corrections are expected to be negligible for sufficiently large black holes (low temperature), but may become relevant for small black holes, where the temperature is higher and the canonical ensemble becomes more sensitive to fluctuations \cite{Das2002,Mustafa2025}.

Retaining only the leading contribution, the corrected entropy can be written in the generic logarithmic form \cite{Das2002}
\begin{equation}
S_c=S-\xi\, \ln \left(S\,T^2\right),
\label{tf1}
\end{equation}
where $S$ is the uncorrected entropy and $T$ is the Hawking temperature. The dimensionless parameter $\xi$ controls the strength of the correction: $\xi=0$ reproduces the classical result, while $\xi\neq 0$ accounts for the leading thermal fluctuation effects. In practice, the correction is suppressed when $T$ is small (large $r_h$), and becomes more important at higher temperatures (small $r_h$).

In the non-extensive setup adopted here, we consider the Tsallis entropy \cite{Tsallis2013,Mustafa2025}
\begin{equation}
S=\delta \left( \pi r_h^2\right)^{\kappa }\qquad (\kappa \geq 1),
\label{tf2} 
\end{equation}
where $\delta$ and $\kappa$ parameterize deviations from the standard area law. Combining Eq.~(\ref{tf2}) with the Hawking temperature in Eq.~(\ref{dd4}) and substituting into Eq.~(\ref{tf1}), we find
\begin{align}
S_c
&=\delta \left( \pi r_h^2\right)^{\kappa}
-\xi \ln\!\left[
\frac{\delta\,\pi^{\kappa-2}\,r_h^{2\kappa-2}}{16}\,
\left\{
\frac{1-k \eta^2}{1-\ell}
-\frac{e^{-\gamma}\,Q^2}{(1-\ell)^2 r^2_h}
+8\pi P\,r^2_h
\right\}^2
\right].
\label{tf3}
\end{align}

On the other hand, if one uses the standard Bekenstein--Hawking entropy given in Eq.~(\ref{dd5}), the corrected entropy becomes
\begin{align}
S_c
=\pi r_h^2
-\xi \ln\!\left[
\frac{1}{16 \pi}\,
\left\{
\frac{1-k \eta^2}{1-\ell}
-\frac{e^{-\gamma}\,Q^2}{(1-\ell)^2 r^2_h}
+8\pi P\,r^2_h
\right\}^2
\right].
\label{tf4}
\end{align}

Equations~(\ref{tf3}) and~(\ref{tf4}) make clear that the leading correction depends on the full temperature structure of the black hole, and therefore inherits the parameter dependence of Eq.~(\ref{dd4}). In particular, $\ell$ and $\eta$ control the geometric prefactor $(1-k\eta^2)/(1-\ell)$, $\gamma$ suppresses the electromagnetic sector through $e^{-\gamma}$, and $P$ governs the AdS contribution at large $r_h$. As a consequence, thermal fluctuations are expected to be most pronounced in the small-$r_h$ regime, where the competition among the geometric, charge, and AdS terms can significantly modify $T$ and hence the logarithmic correction.

To quantify the magnitude and parameter sensitivity of the thermal fluctuation correction, we define the dimensionless ratio
\begin{equation}
\frac{\Delta S}{S}\equiv\frac{S_c-S}{S}=-\frac{\xi\,\ln(S\,T^2)}{S},
\label{tf5}
\end{equation}
which measures the relative departure of the corrected entropy from the uncorrected Bekenstein--Hawking value. For the standard area-law entropy $S=\pi r_h^2$ and the Hawking temperature in Eq.~(\ref{dd4}), Eq.~(\ref{tf5}) reduces to
\begin{equation}
\frac{\Delta S}{S}=-\frac{\xi}{\pi r_h^2}\,\ln\!\left[
\frac{1}{16 \pi}\,
\left\{
\frac{1-k \eta^2}{1-\ell}
-\frac{e^{-\gamma}\,Q^2}{(1-\ell)^2 r^2_h}
+8\pi P\,r^2_h
\right\}^2
\right].
\label{tf6}
\end{equation}
The factor $1/r_h^2$ in front of the logarithm ensures that the correction is suppressed for large black holes ($r_h\gg 1$), as physically expected, while it becomes significant for small horizons. The argument of the logarithm, $S\,T^2$, encodes the full dependence on the geometric and electromagnetic parameters.

\begin{figure*}[htb!]
\centering
\includegraphics[width=\textwidth]{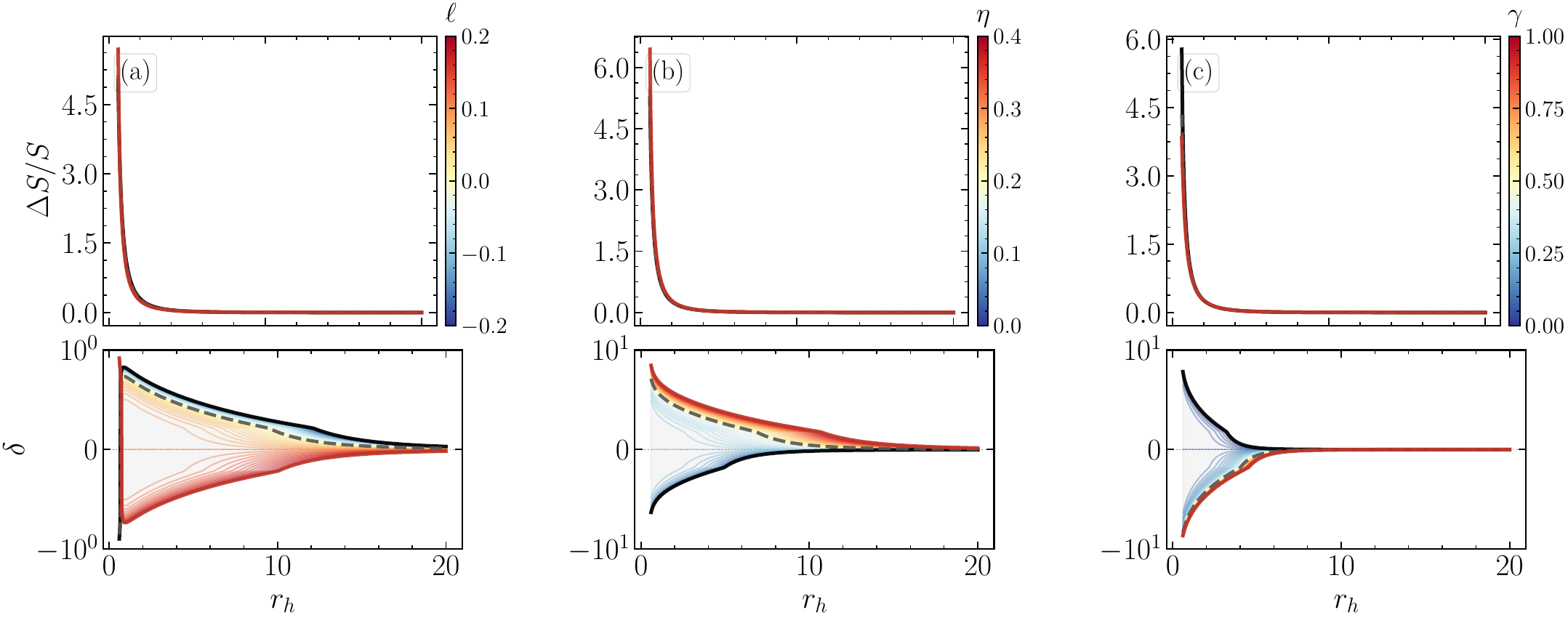}
\caption{Entropy correction ratio $\Delta S/S=(S_c-S)/S$ [Eq.~(\ref{tf6})] as a function of $r_h$ for $\xi=1$, $Q=0.5$, $k=1$, and $P=0.003$, using the Bekenstein--Hawking entropy $S=\pi r_h^2$ and the Hawking temperature from Eq.~(\ref{dd4}). \textbf{Top row:} entropy correction ratio. \textbf{Bottom row:} absolute deviation $\delta=(\Delta S/S)_{\rm param}-(\Delta S/S)_{\rm ref}$ relative to the reference parameter set $(\ell_0,\eta_0,\gamma_0)=(0.1,0.1,0.1)$, displayed on a symmetric logarithmic scale. In each column, one parameter is varied continuously and encoded by the color gradient (see colorbar), while the remaining parameters are fixed at their reference values. Three anchor curves (minimum, reference, and maximum of the swept parameter) are highlighted. Panel~(a): $\ell\in[-0.2,\,0.2]$ (with $\eta=0.1$ and $\gamma=0.1$ fixed). Panel~(b): $\eta\in[0,\,0.4]$ (with $\ell=0.1$ and $\gamma=0.1$ fixed). Panel~(c): $\gamma\in[0,\,1]$ (with $\ell=0.1$ and $\eta=0.1$ fixed). The shaded regions in the residual subpanels indicate the full parameter envelope.}
\label{fig:entropy_corr}
\end{figure*}

In Fig.~\ref{fig:entropy_corr}, we display the entropy correction ratio $\Delta S/S$ defined in Eq.~(\ref{tf6}) as a function of $r_h$, evaluated for $\xi=1$ with the Bekenstein--Hawking entropy $S=\pi r_h^2$ and the Hawking temperature from Eq.~(\ref{dd4}). The top row of panels presents $\Delta S/S$ itself, while the bottom row shows the absolute deviation $\delta=(\Delta S/S)_{\rm param}-(\Delta S/S)_{\rm ref}$ with respect to the reference parameter set $(\ell_0,\eta_0,\gamma_0)=(0.1,0.1,0.1)$, plotted on a symmetric logarithmic scale that captures the wide dynamic range spanning several orders of magnitude. As in the previous figures, the parameter varied in each column is encoded by the color gradient (see the vertical colorbar), with three anchor curves highlighted for guidance.

Across all panels, $\Delta S/S$ is a positive, rapidly decaying function of $r_h$: it attains values of order $\mathcal{O}(5)$ for $r_h\lesssim 1$ and drops below $\mathcal{O}(10^{-2})$ for $r_h\gtrsim 5$. This steep decay reflects the $1/r_h^2$ prefactor in Eq.~(\ref{tf6}), which causes the thermal fluctuation correction to become increasingly negligible as the horizon area grows. For small black holes, by contrast, the correction is substantial and even dominates the classical entropy, signaling that the semiclassical description becomes unreliable in this regime and that a full quantum gravity treatment may be needed.

Figure~\ref{fig:entropy_corr}(a) illustrates the effect of the Lorentz-violating parameter $\ell\in[-0.2,\,0.2]$: the residual $\delta$ spans $\pm\,10^{-1}$ at small $r_h$ and decays below $10^{-4}$ for $r_h\gtrsim 10$. The spread is approximately symmetric about $\delta=0$, reflecting the fact that $\ell$ enters the temperature through the factor $(1-\ell)^{-1}$, which shifts $T$ (and hence $\ln(S\,T^2)$) in opposite directions for positive and negative values of $\ell$. The convergence of the residuals at large $r_h$ confirms that the Lorentz-violating correction to the entropy becomes subdominant relative to the leading area-law term far from the horizon.

Figure~\ref{fig:entropy_corr}(b) shows the influence of the global monopole parameter $\eta\in[0,\,0.4]$: the deviation is predominantly negative for large $\eta$ and reaches $|\delta|\sim\mathcal{O}(1)$ at small $r_h$, indicating that $\eta$ can induce order-unity changes in the corrected entropy near the horizon. Physically, increasing $\eta$ reduces the effective geometric factor $(1-k\eta^2)$ and lowers the Hawking temperature, which in turn decreases the argument $S\,T^2$ of the logarithm. Since $\Delta S/S\propto -\ln(S\,T^2)/S$ and $S\,T^2$ is reduced, the net effect is to increase $\Delta S/S$ for larger $\eta$. The residual subpanel clearly shows the asymmetric, one-sided character of this shift.

figure~\ref{fig:entropy_corr}(c) depicts the role of the ModMax parameter $\gamma\in[0,\,1]$: the pattern is qualitatively similar to that of $\eta$ in the sense that the deviation is predominantly one-sided, with $|\delta|$ reaching $\mathcal{O}(1)$ at small $r_h$ and decaying to $\lesssim 10^{-4}$ at large $r_h$. However, the mechanism differs: increasing $\gamma$ suppresses the negative charge term $-e^{-\gamma}Q^2/[(1-\ell)^2 r_h^2]$ in the temperature bracket, thereby \emph{increasing} $T$ and $S\,T^2$. The resulting enhancement of $\ln(S\,T^2)$ makes the logarithmic correction more negative (i.e., $\Delta S/S$ decreases), which explains the negative residual at intermediate $\gamma$ values. For very large $\gamma$, the charge contribution becomes negligible and the correction saturates to a $\gamma$-independent value set by the geometric and pressure terms alone. This saturation is visible as a compression of the color-coded curves toward a common asymptote in the main panel.

Overall, Fig.~\ref{fig:entropy_corr} demonstrates that the thermal fluctuation correction to the Bekenstein--Hawking entropy is a steeply decreasing function of $r_h$, confirming that the leading logarithmic correction is phenomenologically relevant only for small black holes. The parameter dependence reveals a clear hierarchy: $\eta$ and $\gamma$ produce the largest effects (order-unity changes for $r_h\lesssim 2$), while $\ell$ induces smaller but symmetric perturbations. These results are consistent with the structure of Eqs.~(\ref{tf4}) and~(\ref{tf6}) and highlight that the interplay among the geometric, electromagnetic, and Lorentz-violating sectors is most consequential in the regime where quantum corrections are already significant.

\section{Photon Sphere and Shadow} \label{sec3}

In this section, we analyze the optical properties of the KR-AdS black hole described by the metric (\ref{metric}). For massless photons, the formation of a black hole shadow is of particular interest, as it results from the strong deflection of light near the black hole. Photons with small orbital angular momentum are captured, while those with sufficiently large angular momentum escape, producing a dark region in the observer's sky known as the black hole shadow. Observations of the shadows of M87* and Sagittarius A* by the Event Horizon Telescope (EHT) have opened a new avenue for testing gravity and fundamental physics in the strong-field regime \cite{EHTL1,EHTL4,EHTL6,EHTL12,EHTL15,EHTL17}. Therefore, studying the effects of the bumblebee vector field on black hole shadows provides an opportunity to use observational data to constrain the magnitude of Lorentz-violating parameters.

Considering null geodesic motion in the equatorial plane, defined by $\theta=\pi/2$ and $\dot{\theta}=0$, the Lagrangian density function $\mathcal{L}=\frac{1}{2}\,g_{\mu\nu}\,\dot{x}^{\mu}\,\dot{x}^{\nu}$ \cite{Wald1984} using the metric (\ref{metric}) reduces to:
\begin{equation}
    \mathcal{L}=\frac{1}{2}\,\left[-f(r)\,\dot{t}^2+\frac{\dot{r}^2}{f(r)}+r^2\,\dot{\phi}^2\right].\label{pp1}
\end{equation}
There exist two conserved quantities associated with the cyclic coordinates $(t, \phi)$ given by $\mathrm{E}=f(r)\,\dot{t}$ and $\mathrm{L}=r^2\,\dot{\phi}.$ Eliminating $\dot{t}$ and $\dot{\phi}$ in Eq.~(\ref{pp1}) for null geodesics, we find the radial component motion: $\dot{r}^2=\mathrm{E}^2-V(r)$, where $V(r)$ is the effective potential for null geodesics given by:
\begin{equation}
    V(r)=\frac{\mathrm{L}^2}{r^2}\,f(r).\label{pp2}
\end{equation}

\begin{figure}[tbhp]
\centering
\includegraphics[width=\textwidth]{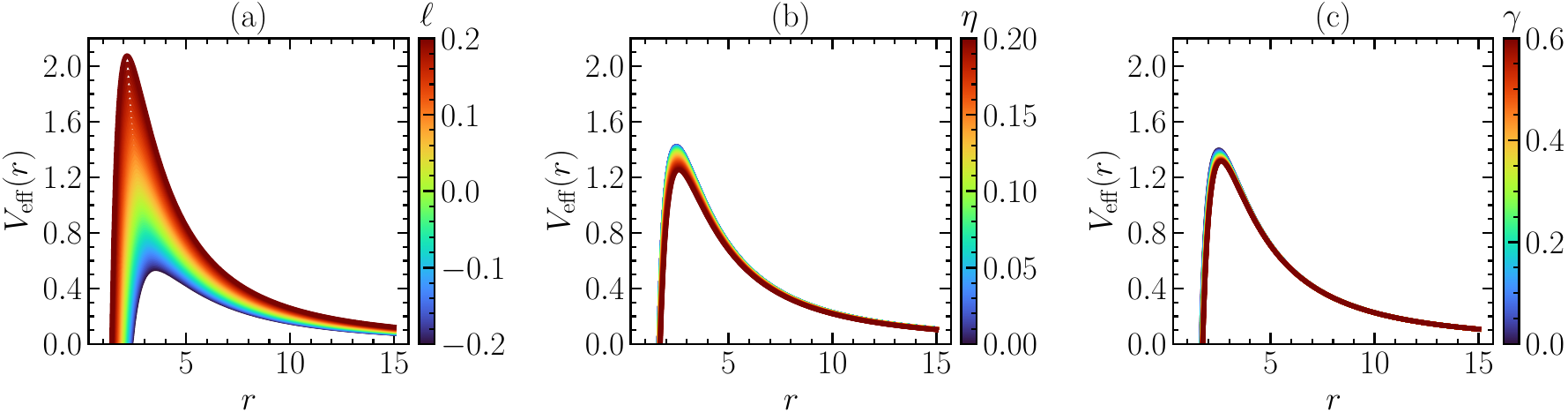}
\caption{Effective potential $V_{\rm eff}(r)$ for null geodesics [Eq.~(\ref{pp2})], for $L=5$, $M=1$, $Q=0.5$, $k=1$, and $P=0$ (asymptotically flat case). In each panel, one parameter is varied continuously and encoded by the color gradient (see colorbar), while the remaining parameters are fixed. Panel~(a): $\ell$ varied continuously (with $\eta=0.1$ and $\gamma=0.1$ fixed). Panel~(b): $\eta$ varied continuously (with $\ell=0.1$ and $\gamma=0.1$ fixed). Panel~(c): $\gamma$ varied continuously (with $\ell=0.1$ and $\eta=0.1$ fixed).}
\label{fig:Veff}
\end{figure}

The effective potential $V(r)$ for null geodesics is displayed in Fig.~\ref{fig:Veff}. In each panel, the potential exhibits a single maximum whose location corresponds to the unstable circular photon orbit (photon sphere). In the updated plots, the parameter varied in each panel is encoded continuously by the color gradient (see the vertical colorbar), which makes the smooth shift of the peak position and height across the parameter interval immediately visible. Panel~(a) shows the effect of varying $\ell$, which modifies the overall scale of $f(r)$ and therefore changes both the height and location of the potential barrier. Panel~(b) shows the analogous deformation under varying $\eta$ through the deficit-angle factor. Panel~(c) shows the effect of varying $\gamma$, which primarily affects the near-peak region because the charge sector enters $f(r)$ through $e^{-\gamma}Q^2/r^2$.

For circular null orbits of radius $r=\text{const.}$, the conditions requiring $\dot{r}=0$ and $\ddot{r}=0$ must be satisfied. These can can be re-written as
\begin{equation}
    \mathrm{E}^2=V(r),\label{pp3}
\end{equation}
and
\begin{equation}
    \frac{d}{dr}\left(\frac{f(r)}{r^2}\right)=0.\label{pp4}
\end{equation}
The photon sphere radius $r_s$ using Eq.~(\ref{pp4}) can be determined as
\begin{equation}
    \frac{1-k \eta^2}{(1 - \ell)}\,r^2 -3\,M\,r + \frac{2 e^{-\gamma} Q^2}{(1 - \ell)^2}=0,\label{pp5}
\end{equation}
with solution:
\begin{equation}
    r_s=\frac{3\,M \,(1-\ell)}{2 (1-k \eta^2)}\,\left[1+\sqrt{1-\frac{8 e^{-\gamma} Q^2}{9 M^2}\frac{1-k \eta^2}{(1-\ell)^3}}\right].\label{pp6}
\end{equation}

Noted that the photon sphere radius is real and finite provided we have a constraint on mass parameter $M$ in terms of charge, ModMax and LV and global monopole charge parameters as $M > \frac{2 e^{-\gamma/2} Q}{3(1-\ell)}\sqrt{\frac{2 (1-k \eta^2)}{1-\ell}}$.

Next, we aim to determine the shadow radius for a static observer located at position $r_O$. Noted that the selected space-time is asymptotically bounded AdS space. Moreover, in the absence of cosmological constant, the lapse function is asymptotically bounded rather than flat space, that is
\begin{equation}
    \lim_{r \to \infty} f(r)=\frac{1-\kappa \eta^2}{1-\ell} \neq 1.
\end{equation}

Therefore, following the methodology and procedure adopted in \cite{Volker2022}, we determine the shadow radius. For a static observer at position $r_O$, the shadow radius is given by \cite{Volker2022}: 
\begin{equation}
R_s=r_s\,\sqrt{\frac{f(r_O)}{f(r_s)}}.\label{pp11}
\end{equation}
In our case at hand, we find the following expression
\begin{equation}
    R_s=r_s\,\sqrt{\frac{\frac{1-k \eta^2}{1 - \ell} - \frac{2\,M}{r_O} +\frac{e^{-\gamma}\,Q^2}{(1 - \ell)^2\,r^2_O}-\frac{\Lambda}{3\,(1-\ell)}\,r^2_O}{\frac{1-k \eta^2}{1 - \ell} - \frac{2\,M}{r_s} +\frac{e^{-\gamma}\,Q^2}{(1 - \ell)^2\,r^2_s}-\frac{\Lambda}{3\,(1-\ell)}\,r^2_s}}.\label{final}
\end{equation}

For ModMax-like black hole in KR-gravity with global monopole, the function $f(r_O)$ approaches to $\frac{1-k \eta^2}{1-\ell}$ for the observer located at infinity. In this case, the shadow radius becomes:
\begin{align}
R_s=\frac{r_s\,\sqrt{\frac{1-k \eta^2}{1-\ell}}}{\sqrt{\frac{1-k \eta^2}{1 - \ell} - \frac{2\,M}{r_s} +\frac{e^{-\gamma} Q^2}{(1 - \ell)^2\,r^2_s}}}=\frac{3\sqrt{3}M}{2\sqrt{2}}\,\frac{\left(1+\sqrt{1-\frac{8 e^{-\gamma} Q^2}{9 M^2}\frac{1-k \eta^2}{(1-\ell)^3}}\right)^2\,\frac{(1-\ell)^{3/2}}{(1-\kappa \eta^2)^{3/2}}}{\sqrt{\frac{1-\ell}{1-\kappa \eta^2}-\frac{2e^{-\gamma} Q^2}{3M^2(1-\ell)^2}+\frac{1-\ell}{1-\kappa \eta^2}\,\sqrt{1-\frac{8 e^{-\gamma} Q^2}{9 M^2}\frac{1-k \eta^2}{(1-\ell)^3}}}},\label{pp7} 
\end{align}
where we have used $r_s$ is given in Eq.~(\ref{pp6}).

In the limit $\eta=0$, corresponding to the absence of global monopole and $\gamma=0$, corresponding to the absence of ModMax parameter, the selected space-time simplifies to the RN-like black hole in KR-gravity. In that case, the shadow radius reduces to
\begin{align}
R_s&= \frac{9M^2}{4}\,\frac{\left(1+\sqrt{1-\dfrac{8 Q^2}{9 M^2}\dfrac{1}{(1-\ell)^3}}\right)^2 \,(1-\ell)^{3/2}}
   {\sqrt{\dfrac{3 M^2 (1-\ell)}{2}- \dfrac{Q^2}{(1-\ell)^2}+ \dfrac{3 M^2 (1-\ell)}{2}\sqrt{1-\dfrac{8 Q^2}{9 M^2}
   \dfrac{1}{(1-\ell)^3}}}},
\label{shado2}
\end{align}
which is similar to the result obtained in \cite{ref2}.

Moreover, in the limit $Q=0$, corresponding to the absence of the electric charge, the shadow radius simplifies as
\begin{equation}
    R_s=3\sqrt{3}\,M\,\frac{1-\ell}{1-\kappa \eta^2}\label{shadow1}
\end{equation}
which is similar to the result obtained in \cite{ref3}.

From the above expressions for photon sphere radius $r_s$ given in Eq.~(\ref{pp6}) and the shadow radius in (\ref{pp7}), we observe that all the geometric parameters $(Q,\, \ell,\, \eta,\, \gamma)$ of the black hole spacetime influences these optical characteristics. Thus, the results get modification in comparison to those known for the standard Schwarzschild or RN black holes.

\begin{figure}[tbhp]
\centering
\includegraphics[width=\textwidth]{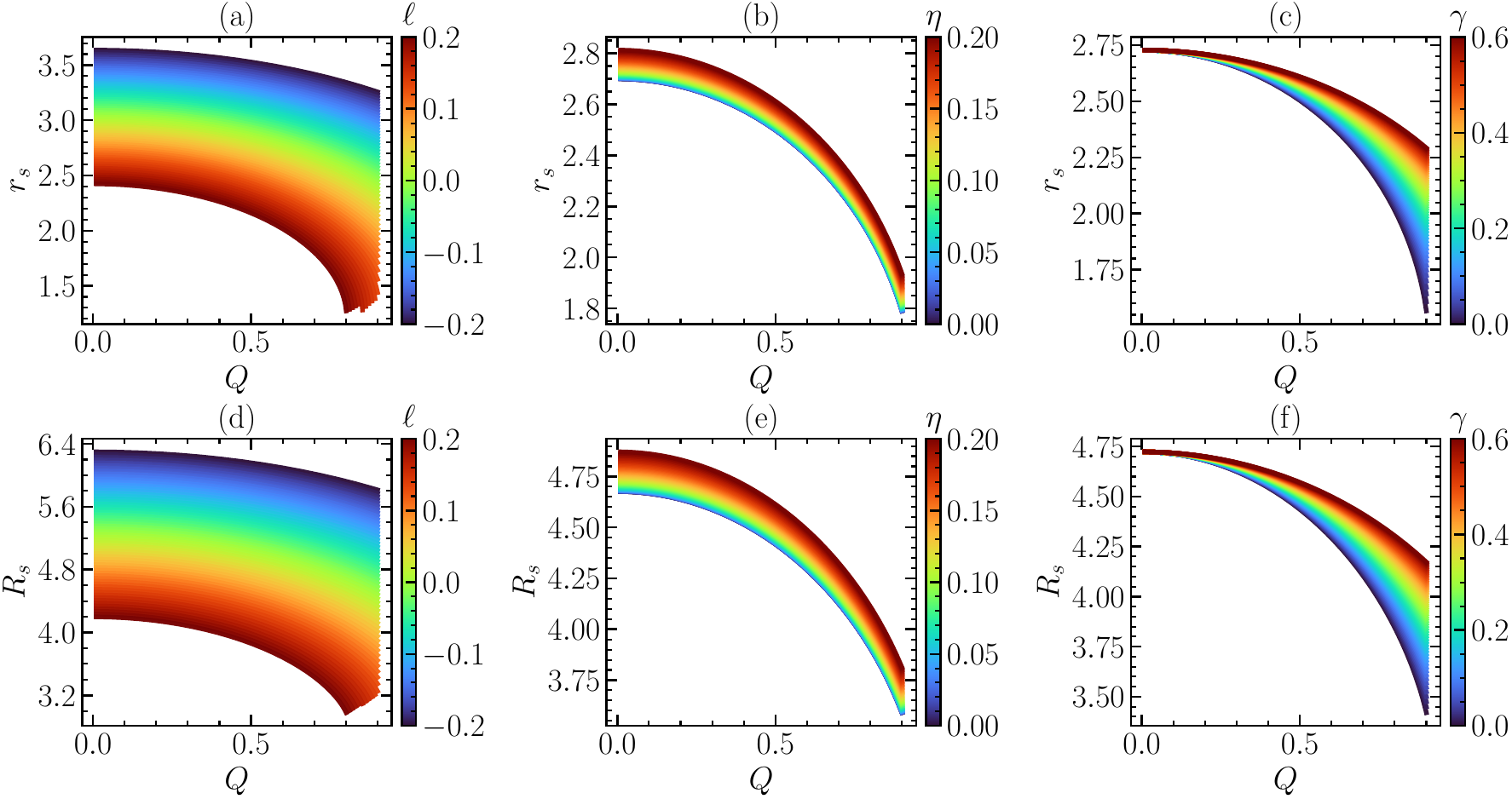}
\caption{Photon-sphere radius $r_s$ (top row) and shadow radius $R_s$ (bottom row) as functions of the electric charge $Q$ [Eqs.~(\ref{pp6}) and~(\ref{pp7})], for $M=1$ and $k=1$. Panels~(a) and (d): $\ell$ varied continuously and encoded by the color gradient (see colorbar), with $\eta=0.1$ and $\gamma=0.1$ fixed. Panels~(b) and (e): $\eta$ varied continuously (color-coded), with $\ell=0.1$ and $\gamma=0.1$ fixed. Panels~(c) and (f): $\gamma$ varied continuously (color-coded), with $\ell=0.1$ and $\eta=0.1$ fixed.}
\label{fig:photon_shadow}
\end{figure}

The photon sphere radius and shadow radius as functions of the electric charge are presented in Fig.~\ref{fig:photon_shadow}. The top row displays the photon sphere radius $r_s(Q)$ [Eq.~(\ref{pp6})], while the bottom row shows the corresponding shadow radius $R_s(Q)$ [Eq.~(\ref{pp7})]. In the updated figure, the parameter varied in each column is encoded by a continuous color gradient (see the corresponding color-bar), so that for each $Q$ one can read off the smooth shift of $r_s$ and $R_s$ as the chosen parameter is tuned across its interval. In panels (a) and (d), varying $\ell$ changes the overall gravitational scaling and therefore shifts both $r_s$ and $R_s$ relative to the reference case. In panels (b) and (e), varying $\eta$ modifies the geometry through the solid-angle deficit and produces a systematic deformation of both radii. In panels (c) and (f), varying $\gamma$ suppresses the effective charge sector through $e^{-\gamma}$, thereby modifying how quickly $r_s(Q)$ and $R_s(Q)$ decrease with $Q$.

We use observational data from the Event Horizon Telescope (EHT) on the black hole shadow of Sagittarius A* (Sgr A*), the supermassive black hole at the center of the Milky Way, to place constraints on the parameters $\gamma$ and $\eta$. The radius of the observed shadow is defined as  
\begin{equation}
R_{\mathrm{SgrA^*}}=\frac{D\,\Theta_s}{2M_{\mathrm{SgrA^*}}},
\label{mm1}
\end{equation}
where $\Theta_s$ denotes the angular diameter of the shadow reported by the EHT,
$M_{\mathrm{SgrA^*}}$ is the mass of Sgr~A*,
and $D$ is its distance from Earth.
Based on combined measurements from the Keck and VLTI instruments, the observational inputs are
$\Theta_{s} = (48.7 \pm 7)\,\mu\mathrm{as}$,
$M_{\mathrm{SgrA^*}} = 4 \times 10^{6}\,M_{\odot}$,
and $D = 8\,\mathrm{kpc}$ \cite{EHTL12,EHTL15,EHTL17}.
Substituting these values into Eq.~\eqref{mm1}, the shadow radius in units of the black hole mass becomes
\begin{equation}
R_{\mathrm{SgrA^*}} = (5.0 \pm 0.5)\, M_{\mathrm{SgrA^*}}.
\label{mm2}
\end{equation}

Moreover, combining the VLTI and Keck estimates yields a refined observational constraint on the dimensionless shadow radius \cite{CQG}:
\begin{equation}
4.55\,M \le R_s \le 5.22\,M.
\end{equation}

\begin{figure}[tbhp]
\centering
\includegraphics[width=\textwidth]{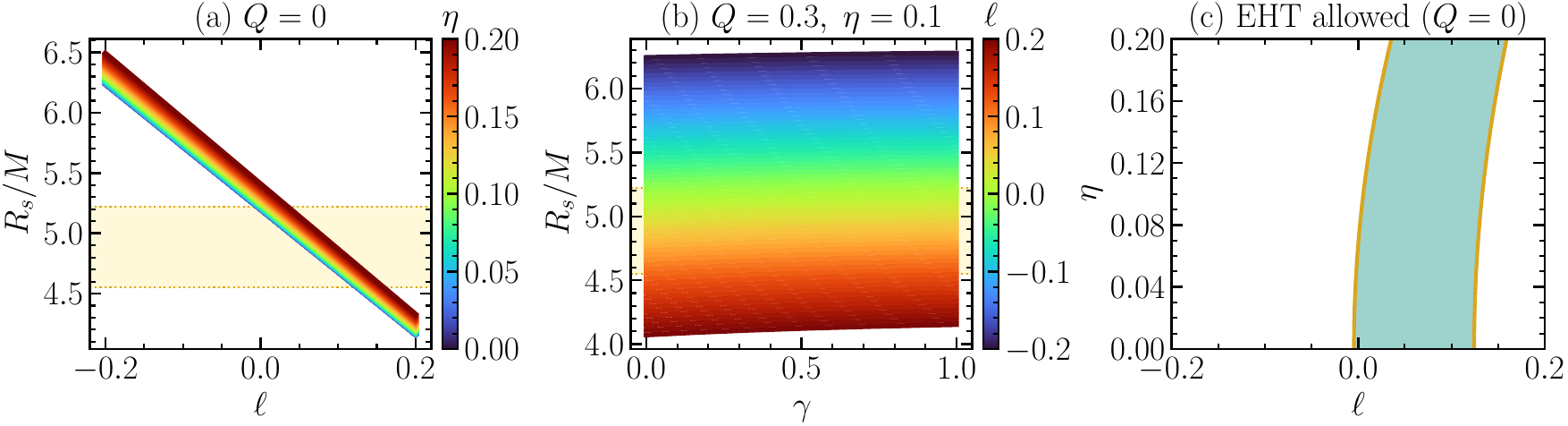}
\caption{EHT observational constraints on the model parameters from the shadow of Sgr~A*. The golden band indicates the observational interval $4.55\,M \leq R_s \leq 5.22\,M$. Panel~(a): $R_s/M$ versus $\ell$ for $Q=0$ (Schwarzschild-like case) with $\eta$ varied continuously and encoded by the color gradient (see colorbar) [Eq.~(\ref{mm1})]. Panel~(b): $R_s/M$ versus $\gamma$ for $Q=0.3$ and $\eta=0.1$, with $\ell$ varied continuously and color-coded [Eq.~(\ref{pp7})]. Panel~(c): allowed region (shaded) in the $(\ell,\eta)$ parameter plane for $Q=0$, bounded by the EHT $1\sigma$ limits.}
\label{fig:EHT_constraint}
\end{figure}

The observational constraints from the EHT shadow measurements of Sgr~A* are presented in Fig.~\ref{fig:EHT_constraint}. The horizontal golden band represents the $1\sigma$ EHT observational window $4.55\,M \leq R_s \leq 5.22\,M$. Panel~(a) displays the Schwarzschild-like limit ($Q=0$) where $R_s/M=3\sqrt{3}(1-\ell)/(1-k\eta^2)$ [Eq.~(\ref{pp7})]. In the updated representation, the dependence on the second parameter is encoded continuously (colorbar), allowing the allowed ranges to be read as a continuous constraint family rather than a set of discrete curves. Panel~(b) shows the charged case ($Q=0.3$, $\eta=0.1$) as $R_s/M$ varies with $\gamma$ for continuously varying $\ell$ (color-coded), emphasizing that $\ell$ is the primary driver of the shift relative to the EHT band while the $\gamma$-dependence is comparatively weak in the displayed range. Panel~(c) maps the allowed region in the $(\ell,\eta)$ parameter space for $Q=0$, with the shaded area representing configurations compatible with the EHT bounds.

Now, we determine the effective radial force experiences by the photons. This force determines the photon capture or escape away by the black hole. The effective force is the negative gradient of the effective potential that governs the photon dynamics. Mathematically, it is defined by
\begin{equation}
    \mathcal{F}=-\frac{1}{2} \frac{\partial V_{\rm eff}}{\partial r}.\label{qq1}
\end{equation}

Using the potential given in Eq.~(\ref{pp2}), we arrive at the following expression:
\begin{equation}
\mathcal{F}=\frac{\mathrm{L}^2}{r^3}\,\left[\frac{1 - k \eta^2}{1 - \ell} - \frac{3 M}{r} + \frac{2 \, e^{-\gamma} Q^2}{(1 - \ell)^2 \, r^2}\right].\label{qq2}
\end{equation}
\begin{figure}[htb!]
\centering
\includegraphics[width=\textwidth]{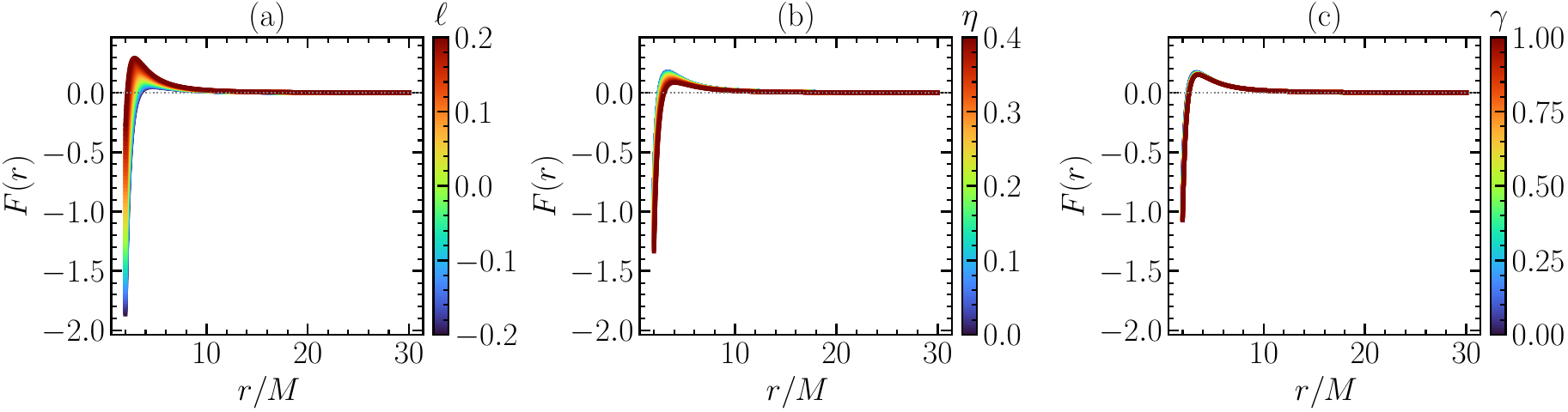}
\caption{Effective radial force $F(r)$ [Eq.~(\ref{qq2})] as a function of the radial coordinate $r/M$ for $M=1$, $Q=0.5$, $k=1$, $\gamma_0=0.1$, $\ell_0=0.1$, $\eta_0=0.1$, and fixed angular-momentum parameter $L=5$. Each panel shows a continuous parameter sweep (color-coded) with a vertical colorbar whose label is placed at the top. Panel~(a): variation of the Lorentz-violating parameter $\ell\in[-0.2,0.2]$ with $\eta=\eta_0$ and $\gamma=\gamma_0$ fixed. Panel~(b): variation of the global monopole parameter $\eta\in[0,0.4]$ with $\ell=\ell_0$ and $\gamma=\gamma_0$ fixed. Panel~(c): variation of the ModMax parameter $\gamma\in[0,1]$ with $\ell=\ell_0$ and $\eta=\eta_0$ fixed. The dotted horizontal line marks $F(r)=0$, separating regions where the effective radial force is outward ($F>0$) from those where it is inward ($F<0$).}
\label{fig:force_panels}
\end{figure}

The effective radial force in Eq.~(\ref{qq2}) provides a compact diagnostic of how the geometric deformations controlled by $(\ell,\eta)$ and the nonlinear electromagnetic sector controlled by $\gamma$ reshape the radial dynamics. In our conventions, the overall prefactor $L^2/r^3$ emphasizes that the force becomes increasingly sensitive to the parameter-dependent bracket at smaller radii, while it decays rapidly at large $r$ as $r^{-3}$. The quantity
\begin{equation}
A(\ell,\eta)=\frac{1-k\eta^2}{1-\ell}
\end{equation}
acts as an effective angular-sector rescaling, whereas the charge contribution is modulated by both $e^{-\gamma}$ and the Lorentz-violating factor $(1-\ell)^{-2}$. Consequently, even when $M$ and $Q$ are fixed, changes in $\ell$, $\eta$, or $\gamma$ can shift the radii at which $F(r)$ changes sign and can alter the strength of the near-horizon force. Figure~\ref{fig:force_panels} displays $F(r)$ in three complementary one-parameter sweeps, using the same continuous-variation visual standard adopted throughout our plots (no legends, colorbar-encoded parameter). The horizontal reference line $F(r)=0$ highlights the radii where the bracketed term vanishes, which can be interpreted as transition points between outward and inward effective radial forcing in this reduced description. Because the charge term scales as $r^{-2}$ inside the bracket, its impact is most pronounced at small radii, while the $-3M/r$ term dominates at intermediate scales and the constant term $A$ controls the large-$r$ asymptote.

In Fig \ref{fig:force_panels}(a), varying the Lorentz-violating parameter $\ell$ modifies both the effective angular-sector factor $A$ and the strength of the charge term through $(1-\ell)^{-2}$. Negative $\ell$ enhances $A$ and generally increases the magnitude of the force at moderate radii, whereas positive $\ell$ tends to suppress the same contribution. This interplay produces a systematic deformation of the force profiles and can shift the location of any zero-crossings of $F(r)$, indicating that Lorentz-violation effects can influence the balance between centrifugal-like and attractive contributions in the strong-field region.

In Fig. \ref{fig:force_panels}(b), increasing the monopole parameter $\eta$ reduces the factor $(1-k\eta^2)$ and therefore decreases $A(\ell,\eta)$ for fixed $\ell$. As a result, the large-$r$ behavior of $F(r)$ is suppressed with increasing $\eta$, and the entire family of curves is shifted accordingly. Physically, this reflects the fact that the global monopole induces a solid-angle deficit that effectively weakens the angular-sector contribution encoded in $A$, thereby modifying the radial forcing experienced by null geodesics.

In fig. \ref{fig:force_panels}(c), varying the ModMax parameter $\gamma$ primarily controls the effective strength of the electromagnetic term via the exponential suppression factor $e^{-\gamma}$. Larger $\gamma$ reduces the charge contribution in the near-horizon region, producing a visible flattening of the small-$r$ force profiles while leaving the large-$r$ tail comparatively less affected (since the charge term becomes subdominant at large radii). This behavior isolates the role of nonlinear electrodynamics in regulating the innermost radial dynamics, which is precisely the regime most relevant to photon-sphere structure and shadow formation.

Overall, Fig.~\ref{fig:force_panels} shows that each parameter induces a distinct and identifiable deformation pattern in $F(r)$: $\ell$ affects both the asymptotic offset and the charge amplification, $\eta$ predominantly rescales the angular-sector contribution through $A$, and $\gamma$ selectively suppresses the charge-driven term at small radii. These trends help anticipate how variations in $(\ell,\eta,\gamma)$ propagate into observable strong-lensing features, such as shifts in the critical impact parameter and changes in the morphology of near-critical photon trajectories.

Finally, we focus into the photon trajectories showing how the geometric parameters later these. The equation orbit is given by
\begin{equation}
    \left(\frac{dr}{d\phi}\right)^2=\frac{\dot r^2}{\dot \phi^2}=\frac{r^4}{\beta^2}-\frac{1-k \eta^2}{1 - \ell}\,r^2+2 M r-\frac{e^{-\gamma}\,Q^2}{(1 - \ell)^2}+\frac{\Lambda}{3\,(1-\ell)}\,r^4.\label{qq3}
\end{equation}
Transforming to a new variable via $r(\phi)=u(\phi)$ and after simplification yields:
\begin{equation}
    \left(\frac{du}{d\phi}\right)^2+\frac{1-k \eta^2}{1 - \ell}\,u^2=\frac{1}{\beta^2}+2 M u^3-\frac{e^{-\gamma}\,Q^2}{(1 - \ell)^2} u^4+\frac{\Lambda}{3\,(1-\ell)}.\label{qq4}
\end{equation}

Differentiating both sides w. r. to $\phi$ results the following second-order differential equation:
\begin{equation}
\frac{d^2u}{d\phi^2}+\frac{1-k \eta^2}{1 - \ell}\,u=3 M u^2-2\frac{e^{-\gamma}\,Q^2}{(1 - \ell)^2} u^3.\label{qq5}
\end{equation}
The above differential equation represents the photons trajectory in the selected black hole within KR-gravity. We observe that various geometric parameters ($\eta,\,\ell,\,\gamma,\,Q$) alter the effective radial force and photons trajectory, and thus, deviates from the standard black hole case. 
\begin{figure}[htb!]
\centering
\includegraphics[width=\textwidth]{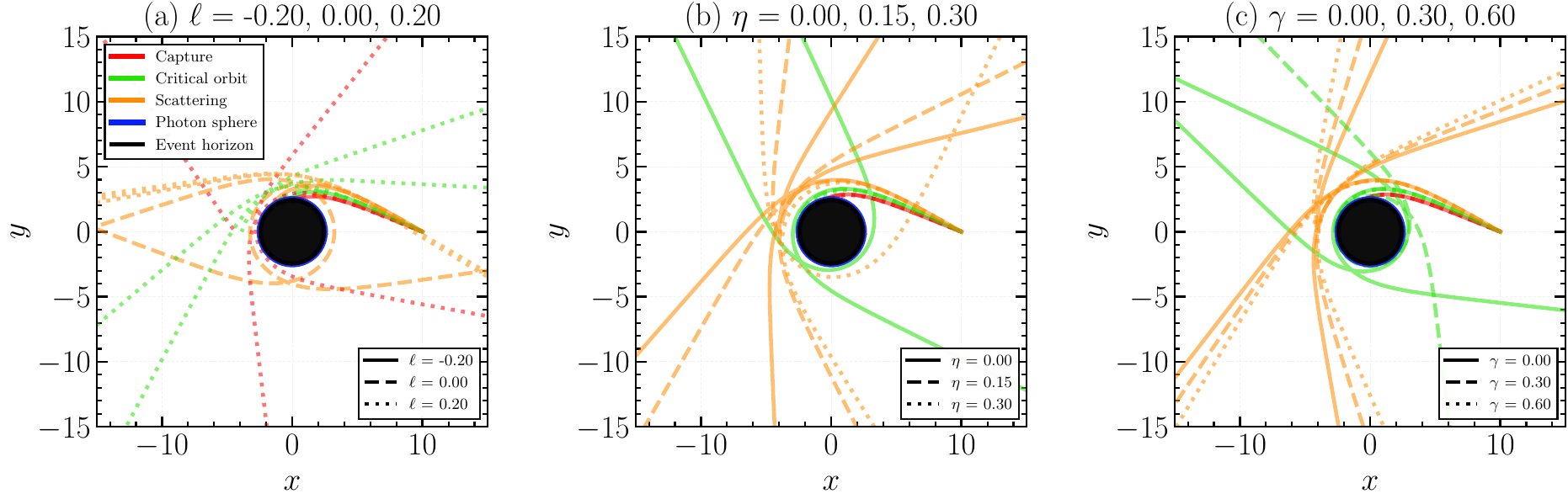}
\caption{Photon trajectories in the $x$--$y$ plane obtained by numerically solving the null geodesic equation [Eq.~(\ref{qq5})] for $M=1$, $Q=0.5$, $k=1$, and $r_0=10$. In each panel, nine trajectories are shown, corresponding to three discrete values of one geometric parameter and three physically distinct trajectory classes. The \emph{trajectory class} is encoded by color: capture (red), critical orbit (green), and scattering (orange), as indicated by the legend in panel~(a). The \emph{parameter value} within each panel is encoded by line style: solid (minimum value), dashed (intermediate value), and dotted (maximum value) of the parameter being varied. The black filled disk represents the event horizon at $r_h \approx 2.5$, while the blue circle marks the photon-sphere reference radius. Panel~(a): $\ell = -0.20,\,0.00,\,0.20$ (with $\eta=0.1$ and $\gamma=0.1$ fixed). Panel~(b): $\eta = 0.00,\,0.15,\,0.30$ (with $\ell=0.1$ and $\gamma=0.1$ fixed). Panel~(c): $\gamma = 0.00,\,0.30,\,0.60$ (with $\ell=0.1$ and $\eta=0.1$ fixed).}
\label{fig:photon_trajectories}
\end{figure}

The photon trajectory equation (\ref{qq5}) reveals that the geometric parameters ($\eta$, $\ell$, $\gamma$) and the charge $Q$ modify the null geodesic dynamics and, consequently, the light-deflection patterns relative to the standard Schwarzschild/RN limits. To visualize these effects, we numerically integrate Eq.~(\ref{qq5}) using representative initial conditions that generate three distinct regimes: capture (photons that cross the horizon), critical orbits (photons that linger near the photon sphere executing multiple loops before eventual capture or escape), and scattering (photons that are deflected and escape to infinity).

Figure~\ref{fig:photon_trajectories} shows the resulting trajectories in Cartesian coordinates for $M=1$, $Q=0.5$, $k=1$, and $r_0=10$. The color coding is the same in all panels and identifies the trajectory class (capture/critical/scattering), whereas the line style changes within each panel to indicate the three discrete values of the parameter being varied (solid/dashed/dotted for minimum/intermediate/maximum). The event horizon is represented by the black disk, and the blue circle provides a reference for the photon-sphere location.

In Fig.~\ref{fig:photon_trajectories}(a), varying the Lorentz-violating parameter $\ell$ in the range $[-0.20,0.20]$ shifts the overall bending pattern and the separatrix between capture and scattering, with the critical trajectories (green curves) showing the strongest sensitivity near the photon-sphere reference circle. Figure~\ref{fig:photon_trajectories}(b) illustrates that increasing the global monopole parameter $\eta$ from $0$ to $0.30$ changes the effective angular sector and modifies both the capture domain and the number of loops performed by near-critical photons. Figure~\ref{fig:photon_trajectories}(c) shows that the ModMax parameter $\gamma$ primarily affects the electromagnetic contribution through the factor $e^{-\gamma}$, which reshapes the innermost portions of the critical and capture trajectories while leaving the large-radius scattering paths comparatively less altered.

The critical trajectories (green curves) trace quasi-spiraling paths around the photon-sphere reference circle, providing a sensitive diagnostic of how the model parameters deform the strong-field geodesic structure. Capture (red curves) characterizes direct infall into the horizon, whereas scattering (orange curves) encodes the net deflection accumulated in the exterior region under the combined influence of $(M,Q)$ and the geometric deformations controlled by $(\ell,\eta,\gamma)$.

Below, we present some well-known results for photon trajectories:
\begin{itemize}
    \item For an electrically charged black hole in KR-gravity \cite{ref2}, the photon trajectories is given by
    \begin{equation}
\frac{d^2u}{d\phi^2}+\frac{1}{1 - \ell}\,u=3 M u^2-\frac{2Q^2}{(1 - \ell)^2} u^3.\label{qq6}
\end{equation}

\item For a neutral black hole with a global monopole in KR-gravity \cite{ref3}, the photon trajectories is given by
\begin{equation}
\frac{d^2u}{d\phi^2}+\frac{1-k \eta^2}{1 - \ell}\,u=3 M u^2.\label{qq7}
\end{equation}

\item For ModMax-AdS black hole solution \cite{DFA2021}, the photon trajectories is given by
\begin{equation}
\frac{d^2u}{d\phi^2}+u=3 M u^2-2 e^{-\gamma}\,Q^2 u^3.\label{qq8}
\end{equation}


\end{itemize}

Thereby, comparing Eq.~(\ref{qq5}) with Eqs.~(\ref{qq6})--(\ref{qq8}), it is clear that  the geometric parameters the electric charge $Q$, the KR-field parameter $\ell$, ModMax parameter $\gamma$, and global monopole parameter $\eta$ changes the photon trajectories in the current case in comparison to the known result in literature.

\section{Conclusion} \label{sec7}

In this work we carried out a detailed thermodynamic and optical analysis of a static, spherically symmetric ModMax--AdS black hole sourced by a global monopole within KR gravity. After constructing the solution and identifying the role played by the geometric parameters (in particular the Lorentz-violating parameter $\ell$, the monopole parameter $\eta$, and the ModMax nonlinearity parameter $\gamma$), we derived the main thermodynamic quantities, including the Hawking temperature, entropy, enthalpy/mass, Gibbs free energy, and the specific heat. We also verified explicitly the consistency of the thermodynamic description by checking the first law and the corresponding Smarr relation, showing that the KR contribution and the geometric deformations are consistently encoded in the extended phase-space framework.

We then investigated the phase structure and criticality in the extended thermodynamics by treating the cosmological constant as a pressure and determining the associated critical points. The resulting phase behavior exhibits a Van der Waals--like structure in the appropriate parameter domain, with the location of the critical point and the coexistence region being significantly shifted by $(\ell,\eta,\gamma)$ (and by the electric charge when present). In the same spirit, we analyzed the Joule--Thomson expansion and obtained the inversion curve, showing that the inversion temperature (including its minimum value) is modified by the space-time parameters, and therefore departs from the standard Maxwell/Einstein-AdS expectations in a controlled way. These results make clear that the KR sector and the ModMax nonlinearity provide additional handles to tune the thermodynamic response of the system.

Finally, we examined the sparsity of Hawking radiation and showed how the emission regime (continuous versus sparse) is affected by the geometric parameters, which effectively reshape the relevant near-horizon scales. On the observational side, we analyzed null geodesics to extract the photon sphere and the shadow radius, demonstrating that the same set of parameters controlling the thermodynamics also leaves an imprint on the optical appearance of the black hole. This provides a direct bridge between the thermodynamic characterization and potentially testable strong-gravity signatures.

As natural extensions of this work, it would be interesting to explore rotating generalizations of the KR-ModMax solution, to analyze quasinormal modes and ringdown constraints, and to combine shadow bounds with thermodynamic consistency requirements to further restrict the allowed parameter space of the model.

\section*{Acknowledgments}

F.A. gratefully acknowledges the Inter University Centre for Astronomy and Astrophysics (IUCAA), Pune, India, for the conferment of a visiting associateship. E. O. Silva acknowledges the support from Conselho Nacional de Desenvolvimento Cient\'{i}fico e Tecnol\'{o}gico (CNPq) (grants 306308/2022-3), Funda\c c\~ao de Amparo \`{a} Pesquisa e ao Desenvolvimento Cient\'{i}fico e Tecnol\'{o}gico do Maranh\~ao (FAPEMA) (grants UNIVERSAL-06395/22), and Coordena\c c\~ao de Aperfei\c coamento de Pessoal de N\'{i}vel Superior (CAPES) - Brazil (Code 001).

\section*{Data Availability Statement}

This manuscript has no associated data.

\end{document}